\input epsf.tex

\documentclass[preprint,pra,aps,showpacs,preprintnumbers,amsmath,amssymb,eqsecnum]{revtex4}

\def\a{{\alpha}}

\def\e{{\epsilon}}

\def\a{\alpha}

\def\au{{\underline \alpha}}

\def\half{\frac{1}{2}}

\newcommand\beq{\begin{equation}}
\newcommand\eeq{\end{equation}}
\newcommand\bea{\begin{eqnarray}}
\newcommand\eea {\end{eqnarray}}

\begin{document}


\title{Arrival Times, Complex Potentials and Decoherent Histories}

\author{J.J.Halliwell}%
\email{j.halliwell@ic.ac.uk}

\author{J.M.Yearsley}
\email{james.yearsley03@imperial.ac.uk}

\affiliation{Blackett Laboratory \\ Imperial College \\ London SW7
2BZ \\ UK }

\date{\today}

\begin{abstract}
We address a number of aspects of the arrival time problem
defined using a complex potential of step function form.
We concentrate on the limit of a weak potential, in which
the resulting arrival time distribution function is
closely related to the quantum-mechanical current.
We first consider the analagous classical arrival time
problem involving an absorbing potential, and this sheds
some light on certain aspects of the quantum case.
In the quantum case, we review the path decomposition expansion (PDX),
in which the propagator is factored across a surface of constant time,
so is very useful for potentials of step function form.
We use the PDX to derive the usual scattering
wave functions and the arrival time distribution function.
This method gives a direct and geometrically
appealing account of known results (but also points the way to how they can
be extended to more general complex potentials). We use these results
to carry out a decoherent histories analysis of the arrival time
problem, taking advantage of a recently demonstrated connection
between pulsed measurements and complex potentials. We obtain very simple
and plausible expressions for the class operators (describing the amplitudes
for crossing the origin during intervals of time) and show that decoherence
of histories is obtained for a wide class of initial states
(such as simple wave packets and superpositions of wave packets). We find that the decoherent
histories approach gives results with a sensible classical limit that are
fully compatible with standard results on the arrival time problem.
We also find some interesting connections between backflow and decoherence.

\end{abstract}

\maketitle

\section{Introduction}

\subsection{The Arrival Time Problem}

The arrival time problem has attracted some considerable interest in recent
years \cite{time}. In the
one-dimensional statement of this problem, one considers an initial
wave function $ | \psi \rangle $ concentrated in the region $x>0$ and consisting
entirely of negative momenta. The question is then to find the
probability $ \Pi (\tau) d \tau $ that the particle crosses $x=0$ between time $ \tau$
and $ \tau + d \tau$. (See Figure 1.)

Two particular candidate expressions for the arrival time
distribution $\Pi (\tau)$ are central to the discussion. First, there is the
current density
\beq
J(\tau) = - \frac { 1 } {2 m } \langle
\psi_f (\tau) | \left( \hat p \delta ( \hat x ) + \delta ( \hat x
) \hat p \right) | \psi_f (\tau) \rangle
\label{1.1}
\eeq
where $ | \psi_f (\tau) \rangle $ is the freely evolved state, which
arises from elementary considerations of the Schr\"odinger
equation. (For convenience we work in units in which $ \hbar = 1$).
This is sensible classically but can be negative in the
quantum case for certain states consisting of superpositions of
different momenta \cite{cur,BrMe,back}. Second, a simple operator
re-ordering of $J(t)$ gives the ``ideal" arrival time distribution
of Kijowski \cite{Kij}
\beq
\Pi_K (\tau) = \frac { 1} { m  }
\langle \psi_f (\tau) ||\hat p|^{1/2} \delta ( \hat x ) |\hat p|^{1/2}
| \psi_f (\tau) \rangle
\label{1.2}
\eeq
which is clearly positive. Both of these distributions are measurable \cite{Muga,HSM}.

\subsection{Complex Potentials}

An interesting question is the extent to which such expressions emerge from
more elaborate measurement or axiomatic schemes. There are many such schemes.
Here, we will focus
on expressions for the arrival time distribution arising from the inclusion
of a complex potential
\beq
V(x) =
- i V_0 \theta (-x)
\label{1.3}
\eeq
in the Schr\"odinger equation. With such a potential, the state
at time $\tau$ is
\beq
| \psi (\tau) \rangle = \exp \left( - i H_0
\tau -  V_0  \theta (-x) \tau \right) | \psi \rangle
\label{1.4}
\eeq
where
$H_0$ is the free Hamiltonian. The idea here is
that the part of the wave packet that reaches the origin during
the time interval $[0, \tau]$ should be absorbed, so that
\beq N( \tau )
= \langle \psi (\tau) | \psi (\tau) \rangle
\label{1.5}
\eeq is the
probability of not crossing $x=0$ during the time interval $[0,\tau]$. The
probability of crossing between $ \tau$ and $\tau + d \tau $ is
then
\beq
\Pi (\tau) = - \frac { d N } { d \tau}
\label{1.6}
\eeq
Complex potentials such as Eq.(\ref{1.3}) were originally considered by Allcock in his seminal
work on arrival time \cite{All} and have subsequently appeared in detector models
\cite{Hal1,Muga}. (See also Ref.\cite{complex,HSM} for further
work with complex potentials).
A recent interesting result of Echanobe et al.
is that under certain conditions, evolution according to
Eq.(\ref{1.4}) is essentially the same as pulsed measurements, in which the
wave function is projected onto $x>0$  at discrete time intervals \cite{Ech}.

For large $V_0$, the wave function defined by the evolution Eq.(\ref{1.4})
undergoes significant reflection, with total reflection
in the ``Zeno limit'', $V_0 \rightarrow \infty$ \cite{Zeno}. Here, we are interested
in the opposite case of small $V_0$, where there is small reflection
and Eq.({\ref{1.6}) can give reasonable expressions for the arrival time
distribution. A number of different authors \cite{All,Muga,HSM} indicate that the
resulting distribution is of the form
\beq
\Pi (\tau) = \int_{-\infty}^{\infty} dt \ R (V_0, \tau - t) J (t)
\label{1.7}
\eeq
where $J(t)$ is the current Eq.(\ref{1.1}) and
\beq
R (V_0, t) = 2 V_0 \theta (t) \exp \left( - 2 V_0 t \right)
\label{1.8}
\eeq
It is therefore closely related to the current $J(t)$ but
also includes the influence of the complex potential
via the ``apparatus resolution function''
$ R(V_0,t)$.

The first aim of this paper is to look in some detail at the calculation
and properties of the arrival time distribution Eq.(\ref{1.7})
defined using a complex potential. In particular, we will use path
integral methods, which in some ways give are more concise and
transparent than previous derivations (and also suggest generalizations
to complex potentials more general than Eq.(\ref{1.3})). We will also explore some
of the properties of the result, Eq.(\ref{1.7}).

The second aim of this paper is to carry out a decoherent histories analysis of the
arrival time problem. This turns out to be closely related to the complex
potentials definition of arrival time and was in fact the original motivation
for exploring complex potentials. In brief, our aim is to see if standard results
for $\Pi (\tau) $, such as  Eq.(\ref{1.1}) or Eq.(\ref{1.7}) have a natural place
in the decoherent histories approach.

\subsection{The Decoherent Histories Approach to the Arrival Time Problem}

We now briefly review the decoherent histories approach to the arrival time
problem \cite{HaZa,MiH,Har,Ya1,YaT,AnSa}. In the decoherent histories
approach \cite{GeH1,GeH2,Gri,Omn,Hal2,DoH,Hal5}, probabilities are assigned to histories via the formula,
\beq
p (\a_1, \a_2, \cdots ) =
{\rm Tr} \left( C_{\au} \rho C_{\au}^{\dag} \right)
\eeq
where the class operator $C_{\au} $ denotes a time-ordered string of projectors $P_{\a}$
interspersed with unitary evolution,
\beq
C_{\au}=
P_{\a_n} e^{ - i H (t_n - t_{n-1}) }
P_{\a_{n-1}} e^{ - i H (t_2 - t_1) } \cdots P_{\a_1}
\eeq
and $ \au $ denotes the string $\a_1, \a_2, \cdots \a_n $. The class operator
satisfies the condition
\beq
\sum_{\au} C_{\au} = e^{ - i H \tau}
\label{C1}
\eeq
where $\tau$ is the total time interval, $ \tau = t_n - t_1 $.
Interference between pairs of histories is measured by the decoherence functional,
\beq
D(\au,\au') = {\rm Tr} \left( C_{\au} \rho C_{\au'}^{\dag} \right)
\eeq
It satisfies the relations,
\bea
D (\au, \au') &=& D^* (\au', \au)
\\
\sum_{\au, \au'} \ D( \au, \au') &=& 1
\\
\sum_{\au} D( \au, \au) &=& \sum_{\au} p (\au)  = 1
\label{D3}
\eea
Of particular interest are sets of histories which satisfy the
condition of decoherence, which is
\beq
D (\au, \au') = 0 \ \ \ \ \ {\rm if} \ \ \ \ \au \ne \au'
\eeq
Decoherence implies the weaker
condition of consistency, which is that $ {\rm Re} D(\au,\au') = 0 $ for $\au \ne \au' $, and
this is equivalent to the requirement that the above
probabilities satisfy the probability sum rules. In most situations
of physical interest, both the real and imaginary parts of $  D(\au,\au')$
vanish for $ \au \ne \au'$, a condition we shall simply call decoherence,
and is related to the existence of records \cite{GeH2,Hal5}.
Decoherence is only approximate in general which raises the question of
how to measure approximate decoherence. The decoherence functional satisfies
the inequality \cite{DoH}
\beq
| D( \au, \au' ) |^2 \le p (\au) p ( \au')
\eeq
This suggests that a sensible measure of approximate decoherence is
\beq
| D( \au, \au' ) |^2  \ \ll \  p (\au) p ( \au')
\label{DA}
\eeq

The approach also permits
other types of class operators which are not simply strings of projections, but
sums of such strings. These are often called inhomogeneous histories \cite{Ish}
and are
relevant to questions involving time in a non-trivial way. For example, for a given
class operator $C_{\au}$, the object $ 1 - C_{\au}$ is also a class operator
but it not equal to a simple string of projections. Unlike homogenous histories, inhomogeneous histories
do not satisfy condition Eq.(\ref{D3}) in general, except when there is decoherence.

A quantity closely related to the probabilities is the quasi-probability
\beq
q(\au) = {\rm Tr} \left( C_{\au} \rho e^{ i H \tau} \right)
\label{1.16}
\eeq
Using Eq.(\ref{C1}), this satisfies
\bea
q( \au ) &=&  \sum_{\au'}  \ {\rm Tr} \left( C_{\au} \rho C_{\au'}^\dag \right)
\nonumber \\
&=& p(\au) + \sum_{\au' \ne \au} D( \au, \au')
\label{1.17}
\eea
(where $\au$ is fixed in the sum).
This means that when there is decoherence, the probabilities for histories $p(\au)$
are equal to the simpler expression $ q(\au)$. (The converse is generally not true
except for in very simple circumstances). Note that $q(\au)$ is not positive
(or even real) in general, but it is positive and real when there is decoherence.
These facts turn out to be
relevant to our analysis of the arrival time problem.

We ultimately seek a decoherent histories account of the arrival time
probability $\Pi (t) dt $, the probability for the particle to arrive
in an infinitesimal interval $[t,t + dt]$. However, for simplicity,
we first consider the simpler problem of computing the probability for
arriving during a finite (possibly large) interval $[0, \tau]$.
We consider an incoming state entirely localized in $x>0$, and we
partition the system's histories into two classes: histories that
either cross or do not cross $x=0$ during the time interval $[0,\tau]$.
We seek class operators $C_{c}$, $C_{nc}$ corresponding to these two
classes of histories.

Some earlier papers on the decoherent histories approach adopted
the following definition of the class operators \cite{Har,MiH,HaZa,Ya1,YaT}.
(This definition is problematic, as we shall see, but sets the stage for the corrected
version we shall use here.)
Let $ P = \theta ( \hat x ) $ denote the projection onto the positive
$x$-axis. To define the class operator for histories which do not cross
$x=0$, we split the time interval into $N$ parts
of size $\epsilon$, and the class operator is defined by
\beq
C_{nc} = \lim_{\epsilon \rightarrow 0} P e^{ - i H \epsilon } P \cdots e^{ - i H \epsilon } P
\label{1.13}
\eeq
where there are $N+1$ projections and $N$ unitary evolution operators, and
$N \rightarrow \infty $ in such a way that $ \tau = N \epsilon$
is constant.
The limit actually yields the so-called restricted propagator,
\beq
C_{nc} = g_r (\tau, 0 )
\eeq
This object is also given by the path integral expression
\beq
\langle x_1 | g_r (\tau, 0 ) | x_0 \rangle = \int_r{\cal D} x  \exp ( i S )
\eeq
where the integral is over all paths from $ x(0) = x_0$ to $x(\tau) = x_1$
that always remain in $x(t) > 0 $.
The class operator for crossing the surface is then defined by
\beq
C_c = e^{ -i H \tau } - C_{nc}
\label{1.15}
\eeq

However, as indicated, there is a problem with this definition. The class operator
Eq.(\ref{1.13}) suffers from the quantum Zeno
effect \cite{Zeno} -- projecting continually in time onto the region $x>0$
prevents the system from leaving it and the probability for
not crossing $x=0$ is unity,
\beq
p_{nc} = {\rm Tr} \left( C_{nc} \rho C_{nc}^{\dag} \right) = 1
\eeq
for any initial state.
This is easily seen from the observation that the restricted propagator
defined by the limit of Eq.(\ref{1.13}) may actually be written,
\beq
g_r (\tau, 0 ) = P \exp \left( - i P H P \tau \right)
\eeq
so is unitary in the Hilbert space of states with support only in $x>0$ \cite{Wall,Sch2}.
Differently put, an incoming wave packet evolving according to
the restricted propagator undergoes total reflection, so never crosses
$x=0$. These results are clearly unphysical and have no sensible classical limit.

The problem here is that the system is monitored too closely to allow
the wave packet to actually cross $x=0$. The resolution is therefore to
relax the monitoring in such a way that something interesting can happen.
There are two obvious ways in which this may be achieved.

The first option is to simply decline to take the limit $\epsilon \rightarrow 0 $
in Eq.(\ref{1.13}), so we define the class operator for not crossing to be,
\beq
C^{\epsilon}_{nc} =P e^{ - i H \epsilon } P \cdots e^{ - i H \epsilon } P
\label{1.18}
\eeq
where again $\tau = N \epsilon$, but $N$ is finite and $\epsilon > 0 $. Clearly if $\epsilon$ is large enough
the system will be monitored sufficiently infrequently to let the wave packet
cross $x=0$ without too much reflection. Studies of the quantum Zeno effect suggest that
the appropriate lower limit on $\epsilon$ is
\beq
\epsilon > \frac {1}  { \Delta H_0 }
\label{1.18b}
\eeq
where $H_0$ is the free Hamiltonian \cite{Zeno}.

The second option is to retain the limit in Eq.(\ref{1.13}), but ``soften'' the projections
to POVMs, that is, to replace $P = \theta (\hat x)$ with a function which is
approximately $1$ for large positive $x$, approximately $0$ for large negative $x$,
and with a smooth transition in between.

What we will do in this paper is adopt a definition of the class operator which
involves elements of both of these options. In particular, we exploit the useful
observation of Echanobe et al \cite{Ech}, which is that the string of projections
interspersed with unitary evolution, Eq.(\ref{1.18})
is approximately equivalent to evolution in the presence of the complex potential
Eq.(\ref{1.3}),
\beq
P e^{ - i H \epsilon } P \cdots e^{ - i H \epsilon } P
\ \approx \ \exp \left( - i H_0 \tau - V_0 \theta (- \hat x ) \tau \right)
\label{1.20}
\eeq
This connection follows from first noting that
\beq
P = \theta (\hat x ) \approx  \exp \left( -  V_0  \theta ( - \hat x ) \epsilon  \right)
\label{1.21}
\eeq
as long as
\beq
V_ 0 \epsilon \gg 1
\label{1.22}
\eeq
We now note that the approximate equivalence
\beq
\exp \left( - i H_0 \epsilon \right)
\ \exp \left( -  V_0  \theta ( - \hat x ) \epsilon  \right)
\approx
\exp \left( - i H_0 \epsilon  -  V_0  \theta ( - \hat x ) \epsilon  \right)
\eeq
will hold as long as
\beq
V_0 \epsilon^2 |\langle [H_0, \theta (-\hat x) ] \rangle| \ll 1
\label{1.24}
\eeq
Echanobe et al. put an upper bound on the left-hand side using the Schr\"odinger-Robertson
inequality and the two inequalities together read
\beq
1 \ll V_0^2 \epsilon^2 \ll \frac {V_0} {\Delta H_0}
\eeq
which can be satisfied as long as
\beq
V_0 \gg \Delta H_0
\eeq
(Note that the Zeno limit restriction Eq.(\ref{1.18b}) is not in fact relevant here
since the quantum Zeno effect goes away when exact projectors are replaced by quasi-projectors,
as in Eq.(\ref{1.21}) \cite{AnSa}).

With a pinch of salt, these arguments indicate that a connection something like
Eq.(\ref{1.22}) will hold.
This equivalence and demonstration of it is somewhat heuristic,
but appears to be backed up by numerical evidence \cite{Ech,Yea}. We will accept
it in what follows, but it would be of interest to obtain a more rigorous proof
of this connection.

This is a very useful result, since the problem of evolution in the presence of the
complex potential Eq.(\ref{1.3}) is straightforward to solve, but the evolution
described by Eq.(\ref{1.18}) could be difficult to solve analytically. In our decoherent
histories analysis of the system, we will therefore define the class operator for not crossing
to be
\beq
C_{nc} = \exp \left( - i H_0 \tau - V_0 \theta (- \hat x ) \tau \right)
\eeq
with the crossing class operator defined by Eq.(\ref{1.15}). These definitions will
be extended to class operators $C_c^k$  for crossing during any one of
a set of small intervals $[t_k,t_{k+1}]$ of size $\Delta$. These are
the class operators we need to give a decoherent histories
account of the origin of $\Pi (\tau)$.

Our main result is that for intervals of size $ \Delta \gg 1 / V_0 $, these class
operators are given approximately by
\bea
C_c^k &=&  e^{ - i H_0 \tau} \int_{t_k}^{t_{k+1}} dt \ \ \frac {(-1)} {2m}
\ \left( \hat p \ \delta (\hat x_t)  \    +
\ \delta (\hat x_t ) \ \hat p \right)
\nonumber \\
&=&  e^{ - i H_0 \tau} \left( \theta ( \hat x (t_k) ) - \theta ( \hat x (t_{k+1} ) ) \right)
\eea
Significantly, the dependence on the complex potential has dropped out entirely. We will
show that there is decoherence for an interesting class of states, and, for such states
the probabilities
are then given by Eq.(\ref{1.16}), which has the form
\beq
q(t_k, t_{k+1} ) = \int_{t_k}^{t_{k+1}} dt \ J(t)
\label{1.38}
\eeq
This, we will show, coincides with the expected result Eq.(\ref{1.7}), when $\Pi (t)$ is
integrated over a range of time much greater than $ 1/V_0$. Hence there is complete
agreement with standard results on the arrival time distribution at sufficiently
coarse-grained timescales. Furthermore, our results
shed some light on the problem of backflow -- we find that the situations when
Eq.(\ref{1.38}) is negative are those in which there is no decoherence in which case
probabilities cannot be assigned.

In another paper, we compute the crossing class operators in a simpler but more heuristic
way, by exploring a semiclassical approximation to Eq.(\ref{1.18}) keeping $\epsilon$ finite.
\cite{HaYe2}. The results are essentially the same.

\subsection{This Paper}

The initial motive and overall aim of this paper are to give a decoherent histories account
of the arrival time question which has a sensible classical limit. As stated, we seek
a decoherent histories analysis account of standard results, such as Eq.(\ref{1.7}).
However, along the way we have
found some new derivations of known results and will explore these areas too.

In Section 2, we consider the classical analogue of the arrival time defined by a complex
potential. This sheds light on the form of the result Eq.(\ref{1.7}) and in particular,
the origin of the resolution function Eq.(\ref{1.8}).
In Section 3, we review the path decomposition expansion (PDX), which is a useful path integral
technique for factoring the propagator across the surface $x=0$, so is very useful for
systems with a step function potential.
In Section 4, we use the PDX to derive the scattering wave functions corresponding to
evolution with a complex step potential. These are of course known, but the PDX gives
a useful and concise derivation of them.
In Section 5, we use the PDX to calculation the arrival time distribution function
Eq.(\ref{1.6}), and very readily obtain the expected result of the form Eq.(\ref{1.7}).
In Section 6 and 7, we use the results of the previous sections to carry out the decoherent histories
analysis, as outlined in the previous subsection. We summarize and conclude in Section 8.

\section{The Classical Case}

Before looking at the quantum arrival time problem, it is
enlightening to look at the corresponding classical arrival
time problem defined using an absorbing potential. This gives
some understanding of the expected form of the result
in the quantum case, Eq.(\ref{1.7}).

We consider a classical phase space distribution $w_t (p,q)$, with initial
value $w_0 (p,q)$ concentrated entirely in $q>0$ with only negative momenta.
The appropriate evolution equation corresponding to the quantum case
Eq.(\ref{1.4}) is
\beq
\frac{ \partial w} { \partial t} = - \frac {p} {m} \frac {\partial w} { \partial
q} - 2 V (q) w
\label{2.1}
\eeq
where  $ V(q) = V_0 \theta (-q) $.
This form may be deduced, for example, by computing the evolution equation of the
Wigner function corresponding to Eq.(\ref{1.4}) and dropping the higher order terms (involving
powers of $\hbar$). Eq.(\ref{2.1}) is readily solved and has solution
\beq
w_\tau (p,q) = \exp \left( - 2 \int_0^\tau ds \ V( q - p s / m ) \right)
\ w_0 (p, q - p \tau / m )
\label{2.2}
\eeq

Following the corresponding steps in the quantum case, the survival probability is
\beq
N( \tau ) = \int_{-\infty}^{\infty} dp \int_{-\infty}^{\infty} dq \ w_\tau (p,q)
\eeq
and the arrival time distribution is
\beq
\Pi (\tau ) = - \frac { d N} {d \tau}
= 2 V_0 \int_{-\infty}^{\infty} dp \int_{-\infty}^0 dq \ w_\tau (p,q)
\eeq
where we have made use of Eq.(\ref{2.1}). This expression is conveniently
rewritten by noting that, again using Eq.(\ref{2.1}), $ \Pi (\tau) $ obeys
the equation
\beq
\frac { d \Pi} { d \tau} + 2 V_0 \Pi =
- 2 V_0 \int_{-\infty}^{\infty} dp \ \frac {p} {m} w_\tau (p,0)
\eeq
This may be solved to yield
\beq
\Pi (\tau) = - 2 V_0 \int_0^\tau dt \ e^{ - 2V_0 (\tau - t) }
\ \int_{-\infty}^{\infty} dp \ \frac {p} {m} w_t (p,0)
\eeq
From Eq.(\ref{2.2}), we see that
\beq
w_t (p,0) = \exp \left( - 2 V_0 \int_0^t ds \ \theta ( p s / m ) \right)
\ w_0 (p,  - p t / m )
\eeq
but since the momenta are all negative the exponential factor makes
no contribution. We thus obtain
\beq
\Pi (\tau) =  2 V_0 \int_0^\tau dt \ e^{ - 2V_0 (\tau - t) } \ J(t)
\label{2.8}
\eeq
where
\beq
J(t) = - \int_{-\infty}^{\infty} dp \ \frac {p} {m} w_0 (p,  - p t / m)
\eeq
The current $J(t)$ is the usual classical arrival time distribution that would have been obtained in the absence of the absorbing potential.

The result Eq.(\ref{2.8}) is very close in form to the expected quantum result,
Eq.(\ref{1.7}), differing only in the range of integration.
The lower limit of $- \infty$ in Eq.(\ref{1.7}) arises as a result of the approximations
used in deriving it and this difference is not significant since we expect
the current to be approximately zero anyway for $ t \le 0$.

The classical result Eq.(\ref{2.8}) helps to understand
the role of the resolution function $R(t)$ in the quantum case Eq.(\ref{1.7}) -- it
in essence describes the classical influence of the absorbing potential
used to model the detector. More precisely, $R(t)$ is actually related to a sort of
coarse graining in time, and this we now demonstrate. Eq.(\ref{2.8}) gives the probability
$ \Pi (\tau ) d \tau $
for arriving during the infinitesimal time interval $ [\tau, \tau + d \tau]$. Supposer we consider the probability for
arriving during a finite time interval, $[ \tau_1, \tau_2]$. This is given by
\bea
p (\tau_2, \tau_1) &=& \int_{\tau_1}^{\tau_2} d \tau \ \Pi (\tau)
\nonumber \\
&=& \int_{\tau_1}^{\tau_2} d \tau \int_0^\tau dt \ 2 V_0  \ e^{ - 2V_0 (\tau - t) } \ J(t)
\label{2.10}
\eea
Rearranging the order of integration and integrating over $\tau$ yields,
\bea
p(\tau_2, \tau_1) &=& \int_0^{\tau_1} dt \int_{\tau_1}^{\tau_2} d \tau \ 2 V_0  \ e^{ - 2V_0 (\tau - t) } \ J(t)
\nonumber \\
&& \ \ \ \ + \int_{\tau_1}^{\tau_2} dt \int_{t}^{\tau_2} d \tau \ 2 V_0  \ e^{ - 2V_0 (\tau - t) } \ J(t)
\\
&=& \int_0^{\tau_1} dt \left( e^{- 2V_0 (\tau_1 - t)} -  e^{- 2V_0 (\tau_2 - t)} \right) \ J(t)
\nonumber \\
&&\ \ \ \ + \int_{\tau_1}^{\tau_2} dt \left( 1 -  e^{- 2V_0 (\tau_2 - t)} \right) \ J(t)
\label{2.12}
\eea
We will see in what follows that $1/V_0$ plays a role as a fundamental short timescale in the problem.
So now suppose we assume that $\tau_1$, $\tau_2$ and $ (\tau_2 -\tau_1) $ are all much greater
than $ 1 / V_0 $. It follows that all the exponential terms may be dropped in Eq.(\ref{2.12})
and we obtain the very simple result,
\beq
p (\tau_2, \tau_1) \approx \int_{\tau_1}^{\tau_2} dt \ J(t)
\label{2.13}
\eeq
That is, all dependence on the resolution function $R(t)$ and the complex potential
parameter $V_0$ completely drops out when we look at probabilities defined on timescales
much greater than $1/V_0$. This result is very relevant to the decoherent histories analysis
considered later where it is natural to look at the arrival time during a finite time interval.

\section{The Path Decomposition Expansion}

In this section we describe some useful path integral techniques.
We wish to evaluate the propagator
\beq
g(x_1, \tau | x_0 ,0 ) = \langle x_1 | \exp  \left( -  i H_0 \tau
-   V_0 \theta (- \hat x) f(\hat x) \tau \right)
| x_0 \rangle
\label{12}
\eeq
for arbitrary $x_1$ and $ x_0 > 0$. This may be calculated using a sum over paths,
\beq
g(x_1, \tau | x_0,0 ) = \int {\cal D} x \exp \left( i S \right)
\label{13}
\eeq
where
\beq
S = \int_0^{\tau} dt \left( \half m \dot x^2 + i V_0 \theta (-x) f(x) \right)
\label{14}
\eeq
and the sum is over all paths $ x(t)$ from $x(0) = x_0$ to $x(\tau) = x_1$.

To deal with the step function form of the potential we need to split off the sections
of the paths lying entirely in $x>0$ or $x<0$. The way to do this is to use
the path decomposition expansion
or PDX \cite{PDX,HaOr,Hal3}.
Consider first paths
from $x_0 > 0 $ to $ x_1 < 0 $.
Each path from $x_0>0$ to $x_1 <0$ will typically cross $x=0$
many times, but all paths have a first crossing, at time $t_1$, say. As a consequence
of this, it is possible to derive the formula,
\beq
g(x_1, \tau | x_0,0 ) = \frac {i } {2m} \int_{0}^{\tau} dt_1
\ g (x_1, \tau | 0, t_1) \frac {\partial g_r } { \partial x} (x,t_1| x_0,0) \big|_{x=0}
\label{PDX1}
\eeq
Here, $g_r (x,t|x_0,0)$ is the restricted propagator given by a
sum over paths of the form (\ref{13}) but with all paths restricted to
$x(t) >0$. It vanishes when either end point is the origin but its derivative
at $x=0$ is non-zero (and in fact the derivative of $g_r$ corresponds to
a sum over all paths in $x>0$ which end on $x=0$ \cite{HaOr}).

It is also useful
to record a PDX formula involving the last crossing time $t_2$,
for $x_0>0$ and $x_1 < 0 $,
\beq
g(x_1, \tau | x_0,0 ) =  - \frac {i  } {2m} \int_{0}^{\tau} dt_2
\ \frac {\partial g_r} {\partial x} (x_1, \tau | x, t_2) \big|_{x=0} \ g (0,t_2|x_0,0)
\label{16}
\eeq
These two formulae may be combined to give a first and last crossing version of the PDX,
\beq
g(x_1, \tau | x_0,0 ) =  \frac {1  } {4m^2} \int_{0}^{\tau} dt_2
\int_0^{t_2} dt_1
\ \frac {\partial g_r} {\partial x} (x_1, \tau | x, t_2) \big|_{x=0} \ g (0,t_2| 0,t_1)
\ \frac {\partial g_r } { \partial x} (x,t_1| x_0,0) \big|_{x=0}
\label{16a}
\eeq
This is clearly very useful for a step potential since the propagator is
decomposed in terms of propagation in $ x<0$ and in $x>0$, essentially
reducing the problem to that of computing the propagator along $x=0$,
$  g (0,t_2| 0,t_1) $. (See Figure 2.)

For paths with $x_0>0$ and $x_1 > 0 $, Eq.(\ref{PDX1}) is modified by the
addition of a term $g_r (x_1,t|x_0,0)$, corresponding to a sum over paths
which never cross $x=0$, so we have
\beq
g(x_1, \tau | x_0,0 ) = \frac {1 } {2m} \int_{0}^{\tau} dt_1
\ g (x_1, \tau | 0, t_1) \frac {\partial g_r } { \partial x} (x,t_1| x_0,0) \big|_{x=0}
+ g_r (x_1,t|x_0,0)
\label{PDX2}
\eeq
Again a further decomposition involving the last crossing, as in Eq.(\ref{16a})
can also be included.

The various elements of these expressions are easily calculated
for a potential of simple step function form $ V(x) = V_0 \theta (-x)$.
The restricted propagator in $x>0$ is
given by the method of images expression
\beq
g_r (x_1, \tau |x_0,0) = \theta (x_1 ) \theta (x_0)
\left( g_f (x_1, \tau |x_0,0) - g_f (-x_1, \tau |x_0,0) \right)
\label{17}
\eeq
where $g_f $ denotes the free particle propagator
\beq
g_f (x_1, \tau |x_0,0) = \left( \frac {m} {2 \pi i \tau } \right)^{1/2}
\ \exp \left( \frac {i m (x_1 - x_0)^2 } { 2  \tau} \right)
\eeq
It follows that
\beq
\frac {\partial g_r } { \partial x} (x,t_1| x_0,0) \big|_{x=0} =
2 \frac {\partial g_f } { \partial x} (0,t_1| x_0,0) \theta (x_0)
\label{18}
\eeq
The restricted
propagator in $x<0$ is given by Eq.(\ref{17}), multiplied by
$ \exp ( - V_0 \tau )$. The only complicated
propagator to calculate is the propagation from the
origin to itself along the edge of the potential, and in
the case $ V(x) = V_0 \theta (-x)$ this is given by \cite{Car},
\beq
g(0,t | 0,0)
= \left( \frac {m} {2 \pi i } \right)^{1/2}
\ \frac { \left(1 - \exp(- V_0 t ) \right) } { V_0 t^{3/2}}
\label{27}
\eeq

Using these results we may write down the full solution to the evolution
with a complex potential, described by Eq.(\ref{12}), for an initial
state $ \psi (x)$ with support only in $x>0$ and with negative momenta.
It has the form,
\beq
\psi (x,\tau) = \theta  (-x) \psi_{tr} (x,\tau) + \theta (x) \left(
\psi_{ref} (x,\tau) + \psi_f (x, \tau) \right)
\label{28a}
\eeq
Here, $ \psi_{tr} $ is the transmitted wave function
and is given by the propagation
of the initial state $ \psi (x)$ using the PDX formulae Eq.(\ref{PDX1}) or Eq.(\ref{16a}).

The remaining part $ \psi_{ref} + \psi_f $ is the wave function
obtained by propagating using the PDX formula for initial and final points both in $x>0$,
Eq.(\ref{PDX2})  (rewritten using Eq.(\ref{16a}) if necessary).
It is appropriate to break this into the two pieces $ \psi_{ref}$, $ \psi_f$ defined as follows:
the reflected wave function $\psi_{ref}$ consists of that obtained using the
first term in Eq.(\ref{PDX2}) together with the reflected part $ - g_f (-x_1, \tau | x_0, 0) $
of the restricted propagator, $g_r$. This definition ensures that $\psi_{ref} \rightarrow 0 $
as the complex potential in $x<0$ goes to zero.
The remaining part, $\psi_f $, is the other
part of the restricted propagator so is simply free propagation in $x>0$. This corresponds
to the part of the incoming wave packet that never reaches $x=0$ during the time
interval $[0, \tau]$. This part clearly goes to zero for large $\tau$.

\section{Solution through Stationary Scattering States}

In this section we use the PDX to derive the standard
representations of the scattering solutions to the Schr\"odinger
equation with the simple complex potential Eq.(\ref{1.3}).
These are known results but this derivation confirms the validity of the PDX
method and also allows a certain heuristic path integral approximation to be tested.
The results will also be useful for the decoherent histories analysis
in Section 6.

The transmitted and reflected wave functions are defined above in Eq.(\ref{28a}).
For large $\tau$, a freely evolving packet moves entirely into $x<0$ so that
the free part, $ \psi_f (x, \tau) $ is zero, leaving just the transmitted
and reflected parts.
Following the above definition, the transmitted wave function is given by
\bea
\psi_{tr} (x, \tau) &=& \frac {1} {m^2} \int_0^\tau ds \int_0^{\tau - s} dv
\ \langle x | \exp \left( - i H_0  s \right) \hat p | 0 \rangle \ e^{ - V_0 s }
\nonumber \\
& \times &
\langle 0 | \exp \left(  - i H v \right)  | 0 \rangle
\ \langle 0 | \hat p \exp \left(  - i H_0 ( \tau - v - s) \right) | \psi \rangle
\label{4.2}
\eea
where $ | 0 \rangle $ denotes the position eigenstate $ | x \rangle $ at $ x = 0$.
Also, we have introduced $ s = \tau - t_1 $ and $ v = t_2 - t_1 $, and
$H = H_0 - i V_0 \theta (-x)  $ is the total Hamiltonian.
The scattering wave functions concern the regime of large $\tau$, so we let the upper
limit of the integration ranges extend to $\infty$.

Writing the initial state as a sum of momentum states $ | p \rangle$, and introducing
$ E = p^2 / 2m $, we have
\bea
\psi_{tr} (x, \tau) &=& \frac {1} {m^2} \int dp \int_0^{\infty} ds
\langle x | \exp \left( - i H_0  s \right) \hat p | 0 \rangle \ e^{ i (E+ i V_0) s }
\nonumber \\
& \times & \int_0^{\infty} dv \ \langle 0 | \exp \left(  - i H v \right)  | 0 \rangle \ e^{ i E v}
\ p \langle 0 | p \rangle \ \ e^{ - i E \tau} \psi (p)
\label{4.3}
\eea
To evaluate the $s$ integral, we use the formula \cite{Sch},
\beq
\int_0^{\infty} ds \ \left( \frac {m} {2 \pi i  s } \right)^{1/2}
\exp \left( i \left[ \lambda s + \frac { m x^2} {2 s} \right] \right)
= \left( \frac { m } { 2 \lambda } \right)^{\half} \ \exp \left( i | x | \sqrt {2 m \lambda} \right)
\label{4.4}
\eeq
from which it follows by differentiation with respect to $x$ and setting $\lambda = E+ i V_0$
that
\beq
\int_0^{\infty} ds
\langle x | \exp \left( - i H_0  s \right) \hat p  | 0 \rangle \ e^{ i (E+ i V_0) s }
= m \exp \left( i | x | [ 2m ( E + i V_0 ) ]^{\half} \right)
\eeq
The $v$ integral may be evaluated using the explicit
expression for the propagator along the edge of the potential, Eq.({\ref{27}),
together with the formula,
\beq
\left( \frac {m} {2 \pi i } \right)^{1/2}
\int_0^{\infty} dv \frac { (1 - e^{ - V_0 v}) } { V_0 v^{3/2} } \ e^{ i E v }
= \frac {\sqrt{2m} } { (E+ i V_0)^{\half} + E^{\half} }
\label{4.5}
\eeq
We thus obtain the result,
\beq
\psi_{tr} (x, \tau) = \int \frac{ dp} { \sqrt{2 \pi} }
\exp \left( - i x [ 2m (E+ i V_0)]^{\half} - i E \tau \right) \ \psi_{tr} (p)
\label{4.6}
\eeq
where
\beq
\psi_{tr} (p) = \frac { 2 } { ( 1 + E^{- \half} ( E + i V_0 )^{\half} ) } \ \psi (p)
\label{4.7}
\eeq
Note that in this final result, it is possible to identify the specific
effects of the different sections of propagation: the propagation
along the edge of the potential corresponds to the coefficient in the
transmission amplitude Eq.(\ref{4.7}) (which is equal to $1$ when $V_0 = 0$),
and the propagation from final crossing to the final point produces
the $V_0$ dependence of the exponent. These observations will be useful below.

The reflected wave function $ \psi_{ref} $ is defined above using
the PDX Eq.(\ref{PDX2}) (rewritten using Eq.(\ref{16a}).
The first term in Eq.(\ref{PDX2}), the crossing part, is the same
as the transmitted case, Eq.(\ref{4.3}), except that $V_0 = 0$ in the last segment
of propagation, from $x=0$ to the final point, and also the sign of $x$ is reversed.
We must also add the effects of the reflection part of the restricted propagator,
and this simply
subtracts the reflection of the incoming wave packet.
The reflected
wave function is therefore given by
\beq
\psi_{ref} (x,\tau) = \int \frac{ dp} { \sqrt{2 \pi} }
\exp \left( i x p  - i E \tau \right) \ \psi_{ref} (p)
\label{4.8}
\eeq
where
\bea
\psi_{ref} (p) &=& \psi_{tr} (p) - \psi (p)
\nonumber \\
&=& \frac { 1 -  E^{- \half} ( E + i V_0 )^{\half} ) } { ( 1 + E^{- \half} ( E + i V_0 )^{\half} ) } \ \psi (p)
\eea

We thus see that the PDX very readily gives the standard stationary wave
functions \cite{All}, without
having to use the usual (somewhat cumbersome) technique of matching eigenfunctions at $x=0$. In fact,
this procedure is in some sense already encoded in the PDX.

We now use these exact results to check the validity of a useful approximation in
the path integral representation of the propagator. In the PDX, Eq.(\ref{PDX1}),
the awkward part to calculate (especially for more general potentials)
is the propagation from  $x=0$ to the final point $x_1 < 0 $. The exact propagator
for this section consists of propagation along the edge of the potential
followed by restricted propagation from $x=0$ to $x_1$, as used in Eq.(\ref{16a}).
However, for sufficiently small $V_0$, one might expect that
in the path integral representation
of the propagator,
the dominant contribution will come from paths in the neighbourhood of the straight line path from $x=0$ to $x<0$. These paths lie almost
entirely in $x<0$, and one might expect that the propagator is therefore given approximately by
\beq
\langle x | \exp \left( - i H s \right) | 0 \rangle \approx
\langle x | \exp \left( - i H_0 s \right) | 0 \rangle
  \exp \left( - V_0 s \right)
\label{4.11}
\eeq
It is not entirely clear
that this is the case, however. On the one hand, the usual semiclassical approximation indicates
that paths close to the straight line paths dominate, but on the other hand, paths in
$x<0$ are suppressed, so maybe the wiggly paths that spend less time in $x<0$ make
a significant contribution. Since this approximation is potentially a useful one, it
is useful to compare with the exact result for the transmitted wave packet calculated
above.

We therefore evaluate the following approximate expression for the transmitted wave function,
\beq
\psi_{tr} (x, \tau) =
- \frac {1} {m} \int_0^{\tau} ds \ \langle x | e^{ - i H_0 s }  | 0 \rangle
\ e^{ - V_0 s }
\ \langle 0 | \hat p e^{ - i H_0 (\tau -s ) } | \psi \rangle
\label{4.12}
\eeq
This is the PDX, Eq.(\ref{PDX1}), in operator form with the semiclassical approximation described above
and we have set $ s = \tau - t_1 $. We now take $ \tau \rightarrow \infty$ in the integration
and evaluate. The key integral is,
\beq
\int_0^{\infty} ds \ \langle x | e^{ - i H_0 s }  | 0 \rangle
\ e^{ i (E  + i V_0 ) s}
=\left( \frac { m } { 2 (E + i V_0) } \right)^{\half} \ \exp \left( - i  x  [2 m (E+iV_0)]^{\half} \right)
\label{4.13}
\eeq
where we have used Eq.(\ref{4.4}) (and recall that $ x < 0 $).
The resulting expression for the transmitted wave function is of the form Eq.(\ref{4.6}), with
\beq
\psi_{tr} (p) = \frac {1} { E^{-\half} (E+ i V_0)^{\half} } \psi (p)
\eeq
This agrees with the exact expression for the transmission coefficient Eq.(\ref{4.7})
only when $V_0 = 0$, with the difference of order $ V_0 / E $ for small $V_0$.
This establishes that the approximation is valid for $ V_0 $ much less than
the energy scale of the initial state.

\section{Calculation of the Arrival Time Distribution}

With the general complex potential Eq.(\ref{1.3}), the arrival time distribution
Eq.(\ref{1.6}) is given by
\beq
\Pi (\tau)
= 2  V_0 \langle \psi_{\tau} | \theta ( - \hat x ) | \psi_{\tau} \rangle
\label{11}
\eeq
where
\bea
| \psi_\tau \rangle &=& \exp \left( - i H \tau \right) | \psi \rangle
\nonumber \\
&=& \exp \left( - i H_0 \tau  - V_0 \theta (-\hat x)  \tau  \right) | \psi \rangle
\eea
(so we use $H = H_0 - i V_0 \theta ( - \hat x )  $ to denote the total
non-Hermitian Hamiltonian).
We are interested in calculating this expression for the case in which $V_0$
is much smaller than the energy scale of the initial state. (The very different limit,
of $V_0 \rightarrow \infty $, the Zeno limit, has been explored elsewhere
\cite{Hal6}).

One way to evaluate Eq.(\ref{11}) is to use the transmitted wave functions,
Eq.(\ref{4.3}). However, we give here instead a different and more direct
derivation using the PDX.
We use the first crossing PDX, Eq.(\ref{PDX1}), which is conveniently rewritten
as the operator expression,
\beq
\langle  x |  \exp ( - i H \tau ) | \psi \rangle
=  -  \frac {1} {m} \int_0^{\tau} dt \ \langle x | \exp ( - i H (\tau -t) )
\ \delta  ( \hat x ) \hat p \ \exp \left( - i H_0 t \right)  | \psi \rangle
\label{19}
\eeq
Now note that the operator $\delta (\hat x) = | 0 \rangle \langle 0 | $ has the simple
property that for any operator $A$
\beq
\delta (\hat x) A \delta (\hat x) = \delta ( \hat x ) \langle 0 | A | 0 \rangle
\label{20}
\eeq
(where, recall, $ | 0 \rangle $ denotes the position eigenstate $ | x \rangle$
at $x=0$).
Inserting Eq.(\ref{19}) in Eq.(\ref{11}), together with the property Eq.(\ref{20})
and the change of variables $ s= \tau - t $, $ s' = \tau - t'$,
yields
\bea
\Pi (\tau) &=& \frac {2 V_0 } {m^2}
\int_0^\tau ds' \int_0^{\tau} ds \int_{-\infty}^0 dx
\nonumber \\
& \times & \langle 0 | \exp \left(  i  H^{\dag} s' \right) | x \rangle \langle x | \exp \left(
- i H s \right) | 0 \rangle
\nonumber \\
& \times & \langle \psi | \exp \left( i H_0 (\tau - s') \right)  \hat p \ \delta (\hat x) \ \hat p \ \exp \left(
 - i H_0 (\tau - s)  \right) | \psi \rangle
\label{5.5}
\eea
We are aiming to show that this coincides with Eq.(\ref{1.7}) with the current
Eq.(\ref{1.1}), and the main challenge is to show how the
$ \hat p  \delta (\hat x)  \hat p $ combination turns into the
combination $ \hat p \delta (\hat x )  + \delta (\hat x ) \hat p $ in the
current Eq.(\ref{1.1}).

Consider first the $x$ integral. Since we are assuming small $V_0$,
we may use the semiclassical approximation Eq.(\ref{4.11}), which
reads
\beq
\langle x | \exp \left( - i H s \right) | 0 \rangle
\ \approx\ \left( \frac {m} {2 \pi i  s } \right)^{1/2}
\ \exp \left( i \frac { m x^2 } { 2  s} - V_0 s \right)
\eeq
The $x$ integral may now be carried out, with the result,
\bea
\Pi (\tau) &=& \frac { V_0 } {m^2}
\int_0^\tau ds' \int_0^{\tau} ds \ \left( \frac {m} {2 \pi i   } \right)^{1/2}
\ \frac{ e^{ - V_0 (s+s') } } { (s-s')^{\half} }
\nonumber \\
& \times & \langle \psi_{\tau} | \exp \left( -i H_0 s' \right)
\hat p \ \delta (\hat x) \ \hat p \ \exp \left(
i H_0 s  \right) | \psi_\tau  \rangle
\label{5.7}
\eea
where $ | \psi_{\tau} \rangle $ denotes the free evolution of the
initial state.

We now carry out one of the time integrals. Note that,
\beq
\int_0^\tau ds' \int_0^\tau ds =
\int_0^\tau  ds' \int_{s'}^\tau ds +
\int_0^\tau ds \int_s^\tau ds'
\eeq
In the first integral, we set $u=s'$, $ v = s-s'$, and in the second
integral we set $ u=s$, $v=s'-s$. We thus obtain
\bea
\Pi (\tau) &=& \frac { V_0 } {m^2} \left( \frac {m} {2 \pi    } \right)^{1/2}
\int_0^\tau du \ e^{ - 2V_0 u } \ \int_0^{\tau - u} dv \ \frac {e^{- V_0 v}} {v^{\half}}
\nonumber \\
& \times &
\left[ \ \frac{1} {i^\half} \langle \psi_{\tau} | \exp \left( -i H_0 u \right) \hat p \ \delta (\hat x) \ \hat p \ \exp \left(
i H_0 (u+v)  \right) | \psi_\tau  \rangle \right.
\nonumber \\
&+&  \left.  \frac{1} {(-i)^\half}\langle \psi_{\tau} | \exp \left( -i H_0 (u+v) \right) \hat p \ \delta (\hat x) \ \hat p \ \exp \left(
i H_0 u  \right) | \psi_\tau  \rangle  \right]
\label{5.9}
\eea
The factors of $ 1/ ( \pm i)^{\half} $ in front of each term come from a careful consideration
of the square root in the free propagator prefactor (and must have this form because
$ \Pi (\tau) $ is real).

We will assume that $ V_0 \tau \gg 1 $, which means that the integrals are concentrated around
$ u = v = 0$. This means that we may take the upper limit of the $v$ integral to be $\infty$,
and it may be carried out, to yield,
\bea
\Pi (\tau) &=& 2 V_0 \int_0^\tau du \ e^{ - 2V_0 u }
\nonumber \\
& \times &
\frac {1} {2m} \langle \psi_{\tau - u} |
\ \hat p \ \delta (\hat x)  \   \Sigma (\hat p)  +  \Sigma^{\dag} ( \hat p)
\ \delta (\hat x ) \ \hat p
\ | \psi_{\tau - u} \rangle
\label{5.10}
\eea
where the operator $\Sigma (\hat p) $ is given by
\beq
\Sigma (\hat p ) = \frac { \hat p } { [ 2 m ( H_0 + i V_0)]^\half}
\eeq
For $V_0$ much less than the energy scale,
\beq
\Sigma (\hat p ) \ \approx \ \frac{ \hat p } { | \hat p | }
\eeq
so $\Sigma (\hat p)$ is simply the sign function of the momentum, which is $ - 1$ in this case,
since the initial state consists entirely of negative momenta.
Finally, writing $ u = \tau - t$, we obtain
\bea
\Pi (\tau) &=&  2 V_0 \int_{0}^\tau dt \ e^{ - 2V_0 (\tau - t) }
\ \frac {(-1)} {2m} \langle \psi_t |
\left( \hat p \ \delta (\hat x)  \    +
\ \delta (\hat x ) \ \hat p
\right) | \psi_t \rangle
\nonumber \\
&=&  2 V_0 \int_{0}^\tau dt \ e^{ - 2V_0 (\tau - t) } \ J(t)
\label{5.13}
\eea
We therefore have precise confirmation of the classical result Eq.(\ref{2.8}),
and also agreement with the expected quantum result, Eq.(\ref{1.7}), modulo
the issues already discussed concerning the range of integration of $t$.

Some comments are in order concerning the positivity of
the result for $\Pi (\tau)$. The expression (\ref{5.10}) is positive because it was derived from the manifestly positive
expression Eq.(\ref{5.5}). (Two approximations were used: the semiclassical approximation
Eq.(\ref{4.11}), and
the condition $V_0 \tau \gg 1$, neither of which affect the positivity of the
result.)

However, to obtain the final result Eq.(\ref{5.13}) we took the limit $V_0 \rightarrow 0 $
in the current part only of Eq.(\ref{5.10}), leaving behind the $V_0$-dependent
term in the exponential part, and the resulting expression is not guaranteed to be
positive. In Eq.(\ref{5.13}}), $J(t)$ is not always positive due
to the backflow effect \cite{cur,BrMe}
and integration
over time does not necessarily remedy the situation. (See Appendix
A for a more thorough discussion of this point). The lack of positivity
for a $\Pi (\tau) $ obtained in this way is not surprising since
taking the limit $V_0 \rightarrow 0 $ in one part of the expression Eq.(\ref{5.10})
only but not the other will not necessarily preserve its property of positivity.
The violation of positivity is generally small, however, so Eq.(\ref{5.13}) may still be a good
approximation to the manifestly positive expression Eq.(\ref{5.10}).

It should also be added that it would be misleading to explore the first
order corrections in $V_0$ in going from Eq.(\ref{5.10}) to Eq.(\ref{5.13}),
since comparable correction terms have already been dropped in using
the semiclassical approximation Eq.(\ref{4.11}).



\section{Decoherent Histories Analysis for a Single Large Time Interval}

We now consider the decoherent histories analysis of this system.
We consider an incoming wave packet approaching the origin
from $x>0$ and ask for the probability of crossing during a given time
interval. We do this in two parts: first in this section, using a large time interval
$[0, \tau]$ and second in the next section, using a set of intervals of arbitrary size interval.

\subsection{Class Operators and Probabilities}

We consider first the following simple question. What is
the probability of crossing or not crossing
during the time interval $ [0, \tau] $?
The class operators for not crossing and crossing are
\bea
C_{nc} &=& \exp \left( - i H_0\tau - V(x) \tau \right)
\label{6.1}\\
C_c &=& \exp\left( - i H_0 \tau \right) - \exp \left( - i H_0 \tau - V(x) \tau \right)
\label{6.2}
\eea
and they satisfy
\beq
C_{nc} + C_c = e^{- i H_0 \tau }
\label{6.3}
\eeq
We are interested in the probabilities for not crossing and crossing,
\bea
p_{nc} (\tau)  &=& {\rm Tr} \left( C_{nc} \rho C_{nc}^{\dag} \right)
\\
p_c (\tau) &=& {\rm Tr} \left( C_{c} \rho C_{c}^{\dag} \right)
\eea
and the off-diagonal term of the decoherence functional,
\bea
D_{c, nc} &=& {\rm Tr} \left( C_{nc} \rho C_{c}^{\dag} \right)
\nonumber \\
&=& {\rm Tr} \left( C_{nc} \rho e^{i H_0 \tau} \right) - p_{nc}
\label{6.6}
\eea
These quantities obey the relation
\beq
p_{nc} + p_c + D_{c,nc} + D^*_{c,nc} = 1
\label{6.5}
\eeq
We look for situations where there is decoherence,
\beq
D_{c, nc} = 0
\eeq
(which is usually only approximate), in which case the probabilities
then sum to $1$,
\beq
p_c (\tau ) + p_{nc} (\tau)  = 1
\label{6.7}
\eeq

It is useful to relate some of these expression
to the standard expressions for arrival time $ \Pi (t)$ defined in Eqs.(\ref{1.5}), (\ref{1.6})
(or Eq.(\ref{11}).
To do this, note that $p_{nc}$ is in fact the same as the survival probability,
$N(\tau)$ defined in Eq.(\ref{1.5}), and that $p_{nc}$ obeys the trivial identity
\beq
p_{nc} (\tau)  = 1 + \int_0^\tau dt \ \frac { d p_{nc} } {dt}
\label{6.10}
\eeq
since $p_{nc} (0) = 1 $. It follows that
\beq
p_{nc} (\tau)  = 1 - \int_0^\tau \ dt \ \Pi (t)
\label{6.11}
\eeq
When there is decoherence, Eq.(\ref{6.7}) holds and we may deduce that
\beq
p_c (\tau) = \int_0^\tau \ dt \ \Pi (t)
\eeq
Hence the decoherent histories analysis is compatible with the standard result, but only
when there is decoherence.

\subsection{Calculation of the Decoherence Functional}

We now give two methods for checking for the decoherence of histories.
The first involves expressing the probabilities and decoherence functional
in terms of the transmitted and reflected waves defined in Eq.(\ref{28a}),
which implies that
\bea
\ C_{nc} | \psi \rangle
&=&  \theta ( - \hat x ) | \psi_{tr} \rangle
+ \theta (\hat x ) \left( | \psi_{ref} \rangle + | \psi_f \rangle \right)
\\
\ C_c | \psi \rangle
&= & \theta ( - \hat x ) \left( | \psi_f \rangle - | \psi_{tr} \rangle \right)
- \theta ( \hat x ) | \psi_{ref} \rangle
\eea
The probabilities and decoherence functional are therefore given by
\bea
p_{nc} &=& \langle \psi_{tr} | \psi_{tr} \rangle + \langle \psi_{ref} | \psi_{ref} \rangle
+  \langle \psi_{ref} | \psi_{f} \rangle
+  \langle \psi_{f} | \psi_{ref} \rangle
+  \langle \psi_{f} | \theta ( \hat x ) |\psi_{f} \rangle
\\
p_c &=& \langle \psi_{tr} | \psi_{tr} \rangle + \langle \psi_{ref} | \psi_{ref} \rangle
- \langle \psi_{tr} | \psi_{f} \rangle
- \langle \psi_{f} | \psi_{tr} \rangle
+ \langle \psi_{f} | \theta ( - \hat x ) |\psi_{f} \rangle
\\
D_{c,nc} &=& \langle \psi_{tr} | \psi_{f} \rangle - \langle \psi_{tr} | \psi_{tr} \rangle
- \langle \psi_{ref} | \psi_{ref} \rangle - \langle \psi_{f} | \psi_{ref} \rangle
\eea
(Here for notational convenience we assume that the definition of $ | \psi_{tr} \rangle
$ includes $ \theta ( - \hat x ) $ and that of $ | \psi_{ref} \rangle $ includes
$ \theta ( \hat x ) $, but the definition of the freely evolving part $ | \psi_f \rangle$
does not include a $\theta$ function).

The magnitude of the off-diagonal
term in the decoherence functional may be estimated from the explicit solution for
the scattering states, Eqs.(\ref{4.6}), (\ref{4.8}). If there is substantial reflection, it is easily
seen that the decoherence functional will not be small. So the interesting regime is the one explored in previous sections, namely
$ V_0 \ll E $ (where $E$ is a typical energy scale). In this regime the reflected
wave functions are of order $ V_0 / E $. Furthermore, one can see from Eq.(\ref{4.6}) that
the difference between $ |\psi_{\tau} \rangle $ and $ | \psi_{tr} \rangle$ are
of order $ V_0 / E$. Therefore, the off-diagonal terms and the probability for
not crossing are of order $V_0 / E $, and the probability for crossing is of
order $1$, up to corrections of order $V_0 / E $. Hence there is decoherence
of histories in the regime $  V_0 \ll E $.

There is a second method of demonstrating decoherence which gives a different picture
and will be useful later. Following the general pattern described in
Eqs.(\ref{1.16}), (\ref{1.17}),
consider the quantity,
\beq
q_{nc} (\tau)  = {\rm Tr} \left( C_{nc} \rho e^{i H_0 \tau} \right)
\label{6.18}
\eeq
From Eq.(\ref{6.6}), we see that the decoherence functional may be written,
\beq
D_{c,nc} = q_{nc} (\tau) - p_{nc} (\tau)
\eeq
This means that $q_{nc} = p_{nc}$ when there is decoherence. Or to put it the
other way round, decoherence of histories may be checked by comparing $q_{nc}$
with $p_{nc}$ and this is what we now do. Recall that $p_{nc}$ is given
by Eqs.(\ref{6.10}), (\ref{6.11}) (which hold in the absence of decoherence).
We may write $q_{nc}$ in a similar form:
\beq
q_{nc} (\tau) = 1 + \int_0^\tau dt \ \frac { d q_{nc} } {dt}
\eeq
The integrand is similar to $\Pi (t)$ defined in Eq.(\ref{11}), so we define
\beq
\tilde \Pi (t) \equiv - \frac { d q_{nc} } {dt}
\eeq
We now have that
\beq
q_{nc} (\tau)  = 1 - \int_0^\tau \ dt \ \tilde \Pi (t)
\eeq
It then follows that the decoherence functional is
\beq
D_{c,nc} =  \int_0^\tau \ dt \left ( \Pi (t) - \tilde \Pi (t) \right)
\eeq

To compute the decoherence functional we need to calculate $\tilde \Pi (t)$, which
is given by
\beq
\tilde \Pi (t) =  V_0 \langle \psi | \ \exp \left( i H_0 t \right)
\ \theta (- \hat x)\  \exp \left( - i H_0 t - V_0 \theta ( - \hat x ) t \right)\  | \psi \rangle
\label{6.21}
\eeq
This is almost the same as $ \Pi (t) $ except
that the exponential on the left involves only $H_0$ and not the complex potential
(and also an overall factor of $2$). We therefore follow the calculation of $\Pi (t)$ in Section 5
with small modifications. With a little care, one may see that the
final result is the same as that for $\Pi (t)$, Eq.(\ref{5.13}),
except that $2 V_0$ is replaced with $V_0$, that is,
\beq
\tilde \Pi (t) =  V_0 \int_{0}^t ds \ e^{ - V_0 (\tau - s) } \ J(s)
\eeq
This result holds for timescales greater than $ 1/ V_0$ and under the semiclassical
approximation Eq.(\ref{4.11}) (which required $ E \gg V_0$ so is equivalent to the
requirement of negligible reflection encountered above).
Finally, a calculation similar to that of Eqs.(\ref{2.10})-(\ref{2.13}) implies that
\beq
\int_0^{\tau} dt \ \Pi (t ) \ \approx \ \int_0^{\tau} dt \ J (t)
\eeq
as long as $ V_0 \tau \gg 1 $. Since this result is independent of $V_0$,
$ \tilde \Pi (t)$ will satisfy the same relation. We thus deduce that
\beq
D_{c,nc} \approx 0
\eeq
hence there is decoherence, under the above conditions.

\section{Decoherent Histories Analysis for an Arbitrary Set of Time Intervals}

We now turn to the more complicated question of much more refined histories,
that may cross the origin at any one of a large number of times, during the time interval
$ [ 0, \tau] $. This corresponds more directly to the standard crossing probability,
$ \Pi (t) dt $, the probability that the particle crosses during an infinitesimal
time interval $ [t, t+ dt] $.

\subsection{Class Operators}

We have defined class operators Eqs.(\ref{6.1}), (\ref{6.2})
describing crossing or not crossing during a time interval $[0,\tau]$.
We now split this time interval into $n$ equal parts of size $\e$,
so $ \tau = n \e $ and we seek class operators describing crossing
or not crossing during any one of the $n$ intervals. We first note
that
\beq
e^{ - i H_0 \e } = C_{nc} (\e ) + C_c ({\e} )
\eeq
where $C_{nc} (\e) $ and $C_c (\e )$ are defined as in Eqs.(\ref{6.1}), (\ref{6.2})
except that here they are for a time interval $ [0,\e]$. We now use this to decompose $ e^{ - i H_0 \tau}$
into the desired class operators. We have
\bea
e^{ - i H_0 \tau } &=& \left( e^{ - i H_0 \e } \right)^n
\nonumber \\
&=& \left( e^{ - i H_0 \e } \right)^{n-1} \ \left( C_{nc} (\e ) + C_c ({\e}) \right)
\nonumber \\
&=&
\left( e^{ - i H_0 \e } \right)^{n-1} C_{nc} (\e) + e^{ - i H_0 (\tau - \e) } C_c (\e )
\eea
Repeating the same steps on the first term, this yields,
\beq
e^{ - i H_0 \tau }  = \left( e^{ - i H_0 \e } \right)^{n-2} C_{nc} (2 \e)
+ e^{ - i H_0 (\tau - 2\e) } C_c (\e ) C_{nc} (\e)
+ e^{ - i H_0 (\tau - \e) } C_c (\e )
\eeq
Repeating more times eventually yields,
\beq
e^{- i H_0 \tau} = C_{nc} (\tau ) + \sum_{k=0}^{n-1} \ e^{ - i H_0 (\tau - (k+1) \e ) }
C_c (\e) C_{nc} (k \e )
\label{7.4}
\eeq
From this expression, we see that the class operator for crossing $x=0$ for the first
time during the time interval $ [k \e, (k+1) \e ] $ is given by the summand of the second term,
\beq
C_c ( (k+1) \e, k \e) = e^{ - i H_0 (\tau - (k+1) \e ) }
C_c (\e) C_{nc} (k \e )
\label{7.5}
\eeq

We will not in fact work with the class operator Eq.(\ref{7.5}),
since a more useful similar but alternative expression can also be found. Taking the continuum
limit of Eq.(\ref{7.4}) (and inserting the explicit expression for $C_{nc})$, we obtain,
\beq
e^{- i H_0 \tau} = e^{- i H_0 \tau - V \tau }
+ \int_0^\tau dt \ e^{- i H_0 (\tau - t)} V e^{- i H_0 t - V t}
\label{7.6}
\eeq
(where, recall, $V = V_0 \theta ( - \hat x ) $). This indicates that the class operator
for first crossing during the infinitesimal time interval $ [t, t+dt]$, is
\beq
C_c (t) = e^{- i H_0 (\tau - t)} V e^{- i H_0 t - V t}
\label{7.7}
\eeq
We do not, however, expect histories characterized by such precise crossing time to be decoherent,
so it is natural to consider coarser-grained class operators,
\beq
C_c^k  = \int_{t_k}^{t_{k+1}} dt \ C_c (t)
\label{7.8}
\eeq
which represents crossing during one of the $N$ time intervals $ [t_k, t_{k+1}]$ of size $\Delta$, where
$ t_k = k \Delta $ with $ k = 0,1 \cdots N-1 $ and $ \tau = N \Delta $.
The complete set of class operators $C_{\a}$ for crossing
and not crossing is the set of $N+1$ operators
\beq
C_{\a} = \{ C_{nc}, C_c^k \}
\eeq
and Eq.(\ref{7.6}) implies that they satisfy
\beq
e^{- i H_0 \tau} = C_{nc} + \sum_{k=0}^{N-1} C_c^k
\label{7.10}
\eeq
To check for decoherence of histories we need to calculate two types of decoherence functional
\bea
D_{k k'} &=& {\rm Tr} \left( C_c^k \rho ( C_c^{k'} )^{\dag} \right)
\label{7.11} \\
D_{ k, nc} &=& {\rm Tr} \left( C_c^k \rho ( C_{nc} )^{\dag} \right)
\label{7.12}
\eea
and this will be carried out below.


\subsection{An Important Simplification of the Class Operator}

There turns out to be a very useful simplification in the class operator Eq.(\ref{7.7}). Consider
the amplitude
\beq
\langle x | e^{ i H_0 \tau} C_c (t) | \psi \rangle = V_0 \langle x | e^{- i H_0 t } \theta (- \hat x)
e^{- i H_0 t - V t} | \psi \rangle
\label{7.13}
\eeq
for any $ x $. The right-hand side is very similar to Eq.(\ref{11}), except that there is no complex potential
in one of the exponential terms and also the ``final" state is $ | x \rangle $ not $ | \psi \rangle$.
(And there is also an overall factor of $2$ different). Despite these differences, we may once again
make use of the details of the calculation of Section 5,
and we deduce from the analogous result Eq.(\ref{5.13}),  that
\beq
\langle x | e^{ i H_0 \tau} C_c (t) | \psi \rangle =
V_0 \int_{0}^t ds \ e^{ - V_0 (t - s) }
\ \frac {(-1)} {2m} \langle x |
\left( \hat p \ \delta (\hat x_s)  \    +
\ \delta (\hat x_s ) \ \hat p
\right) | \psi \rangle
\eeq
Like the derivation of Eq.(\ref{5.13}), this is valid under the conditions that all energy
scales are much greater than $V_0$ and all time scales much greater than $ 1 / V_0 $.
Now we integrate this over time to obtain the coarse-grained crossing time operator, Eq.(\ref{7.8}),
and again use approximations of the form Eqs.(\ref{2.10})-(\ref{2.13}) (again using the assumption
of timescales much greater than $1/V_0$), to yield, the remarkably
simple and appealing form,
\beq
e^{ i H_0 \tau} C_c^k = \int_{t_k}^{t_{k+1}} dt \ \ \frac {(-1)} {2m}
\ \left( \hat p \ \delta (\hat x_t)  \    +
\ \delta (\hat x_t ) \ \hat p \right)
\label{7.15}
\eeq
This may also be written even more simply
\beq
e^{ i H_0 \tau} C_c^k = \theta ( \hat x (t_k) ) - \theta ( \hat x (t_{k+1} ) )
\label{7.16}
\eeq

\subsection{Probabilities for Crossing}

The above expressions for the crossing time class operator are the most important results of the paper
and provide an immediate connection to the standard expression for the arrival time distribution.
Supposing for the moment that there is decoherence of histories, we may assign
probabilities to the histories.
The probability for crossing during the time interval $ [t_k, t_{k+1}]$ is
\beq
p (t_k, t_{k+1} ) = {\rm Tr} \left( C_c^k \rho (C_c^k)^{\dag} \right)
\eeq
However, as noted in Eqs.(\ref{1.16}), (\ref{1.17}) when there is decoherence of histories, this expression for
the probabilities for histories is equal to the simpler expression
\bea
q (t_k, t_{k+1} )  &=& {\rm Tr} \left( C_c^k \rho e^{i H_0 \tau } \right)
\nonumber \\
&=&
\int_{t_k}^{t_{k+1}} dt \ \ \frac {(-1)} {2m}
\ \langle \psi |  \left( \hat p \ \delta (\hat x_t)  \    +
\ \delta (\hat x_t ) \ \hat p \right) | \psi \rangle
\nonumber \\
&=& \int_{t_k}^{t_{k+1}} dt \ J(t)
\label{7.18}
\eea
which is precisely the standard result! The expression for the probability $ q (t_k, t_{k+1} )$ is not
positive in general (although is real in this case, as it happens), but when there is decoherence,
it is equal to $ p (t_k, t_{k+1} ) $, which {\it is} positive. Hence the decoherent histories result
coincides with the standard result under the somewhat special conditions of decoherence of histories.

\subsection{Decoherence of Histories and the Backflow Problem}

There is an interesting connection between decoherence of histories and backflow.
To see this, consider the following simple case. We consider histories which either
cross or do not cross the origin during the time interval $[t_1, t_2]$. So the crossing
and not crossing class operators are $C$ and $ 1 - C $, where
\beq
C =  \theta ( \hat x_1 ) - \theta ( \hat x_2 )
\eeq
where we have adopted the notation $ \hat x_k = \hat x (t_k) $
(and for convenience we have dropped the exponential factor which is just a matter
of definition and drops out of all expression of interest).
The decoherence functional is
\bea
D &=& \langle C ( 1 - C) \rangle
\nonumber \\
&=& \langle C \rangle - \langle C^2 \rangle
\label{7.25}
\eea
This may also be written
\beq
D =  - \langle \left( \theta ( - \hat x_1 ) \theta ( \hat x_2 ) + \theta ( \hat x_2 )
\theta ( - \hat x_1 ) \right) \rangle
\label{7.21}
\eeq
a form we will use below to check decoherence.
When there is decoherence, $D = 0 $ and the probability for crossing is
\beq
p (t_1, t_2) = \langle C^2 \rangle = \langle C \rangle
\eeq
As noted above, $ \langle C \rangle $ is the standard result, Eq.(\ref{7.18}),
for the probability of crossing.

There is an interesting connection here between backflow and decoherence. If there is
decoherence, $D$ is zero so
$ \langle C \rangle$ must cancel  $ \langle C^2 \rangle$ in Eq.(\ref{7.25}),
which means that $ \langle C \rangle \ge 0 $, so there is no backflow. Or we
may make a logically equivalent statement:
if there is backflow, $ \langle C \rangle < 0 $, then there cannot be decoherence, since $|D|$ is then
greater than the probability $ \langle C^2 \rangle $. Hence, states with backflow do not permit
decoherence of histories.
(Note that absence of backflow, $ \langle C \rangle \ge 0 $,
is not itself enough to guarantee decoherence -- the stronger condition $D = 0$
must be satisfied).

This is an important result. The quantity $ \langle C \rangle $ is regarded as the
``standard'' result for crossing time probability and its possible negativity is
disturbing. Here, the decoherent histories approach sheds new light on this issue.
In the decoherence histories apporach,
the true probability for crossing is the manifestly positive quantity $ \langle C^2 \rangle$
and this is equal to $ \langle C \rangle $ only when there is decoherence. In particular,
when there is significant backflow, there cannot be decoherence, so probabilities cannot
be assigned and $ \langle C^2 \rangle$ is not equal to $ \langle C \rangle $.

\subsection{The Decoherence Conditions}

The crossing probabilities described above are only valid when all components of the
decoherence functional, Eqs.(\ref{7.11}), (\ref{7.12}), are zero. We therefore
address the issue of finding those states for which there is negligible decoherence.

We consider first the simpler case, of the decoherence functional Eq.(\ref{7.12}). The non-crossing class operator
$C_{nc}$ is given in general by Eq.(\ref{6.1}). However, it simplifies considerably
in the approximations used to derive Eq.(\ref{7.16}), which we adopt here. In particular,
Eq.(\ref{7.10}) with Eq.(\ref{7.16}) imply that
\beq
e^{ i H_0 \tau} = C_{nc} + e^{ i H_0 \tau} \left[ \theta ( \hat x ) - \theta ( \hat x ( \tau) ) \right]
\eeq
Since we are interested only in initial states with support entirely in $x>0$, we have
$ \theta ( \hat x ) | \psi \rangle = | \psi \rangle $, which means that effectively,
\beq
C_{nc} (\tau) \ \approx \  \theta ( \hat x ) e^{ - i H_0 \tau }
\eeq
The decoherence functional of interest is then
\beq
D_{k, nc} = \langle \psi | \theta ( \hat x (\tau) )
 \left[ \theta ( \hat x_k ) - \theta ( \hat x_{k+1}  )\right]
| \psi \rangle
\eeq
This is conveniently
rewritten,
\beq
D_{k, nc} = \langle \psi | \theta ( \hat x (\tau) )
 \left[ \theta ( - \hat x_{k+1} ) - \theta ( -\hat x_{k}  )\right]
| \psi \rangle
\label{7.26}
\eeq
We will take $ \tau $ to be very large and it is pretty clear that this object
will be approximately zero, since we expect all the initial state to end up in
$ x<0$ at large times. However, we will see this below in more detail.

The more important decoherence condition is that $D_{kk'}$ defined  in Eq.(\ref{7.11})
vanishes, so we now focus on that.
We write the class operator Eq.(\ref{7.16}) for crossing during the time
interval $ [t_k, t_{k+1}]$ as
\beq
C_c^k = e^{ - i H_0 \tau} \left( \theta ( \hat x_k ) - \theta ( \hat x_{k+1}  ) \right)
\eeq
For an initial state $ | \psi \rangle$, the quantity
$ C_c^k | \psi \rangle $ is a quantum state representing the property of
crossing of the origin in the time interval $ [t_k, t_{k+1}]$. The decoherence
condition $D_{kk'} = 0$ is simply the condition that the ``crossing states''
$ C_c^k | \psi \rangle $ for different
time intervals have negligible interference.
The states
$ C_c^k | \psi \rangle $ consist of an initial state
which has been localized to a range of time at $x=0$.
This is closely related to the interesting question
of diffraction in time \cite{diff} and this connection will be explored in more
detail elsewhere \cite{HaYe}.

The decoherence functional is given by
\bea
D_{kj} &=& \langle \left( \theta ( \hat x_k ) - \theta ( \hat x_{k+1}  ) \right)
\left( \theta ( \hat x_{j} ) - \theta ( \hat x_{j+1}  ) \right) \rangle
\nonumber \\
&=& \langle \left( \theta ( \hat x_k ) - \theta ( \hat x_{k+1}  ) \right)
\left( \theta ( -\hat x_{j+1} ) - \theta (- \hat x_{j}  ) \right) \rangle
\eea
where without loss of generality we take $ t_{j+1} < t_k $.
It is a sum of terms each of the form,
\beq
d_{kj} =  \langle \theta ( - \hat x_k ) \theta ( \hat x_j )  \rangle
\eeq
where $t_k < t_j $. Note that
\beq
| d_{kj} |^2 \le  d^2_m = \langle  \theta ( - \hat x_k ) \theta ( \hat x_j )  \theta ( - \hat x_k ) \rangle
\label{7.30}
\eeq
The key thing is that $d_m^2$ has the form of a probability -- it is the probability to
find the particle in $x<0$ at $t_k$ and then in $x>0$ at $t_j $. Semiclassical
expectations suggest that this is small in general for the states considered here,
which are left-moving wave packets, and indeed the classical limit of this probability
is zero. So this is a useful object to calculate in terms
of checking decoherence. (Although it may not be small for states with
backflow). Note that it also implies that the other parts of the decoherence functional,
Eq.(\ref{7.26}) will also be small. In detailed calculations, the upper bound
Eq.(\ref{7.30}) must be compared with the probabilities, as in Eq.(\ref{DA}).

\subsection{Checking the Decoherence Condition for Wave Packets}

We now consider the particular case of an initial state consisting of a wave packet
\beq
\psi (x) = \frac{1} {(2 \pi \sigma^2)^{1/4} } \exp \left( - \frac {(x-q_0)^2} {4 \sigma^2 } + i p_0 x \right)
\label{7.33}
\eeq
where $q_0 > 0 $ and $ p_0 < 0 $. We first consider a heuristic analysis of decoherence.
In the simplest case, the wave packet crosses
the origin almost entirely during the time interval $ [t_k, t_{k+1}]$ (of size $\Delta$)
for some $k$, without
any substantial overlap with any other time intervals. (See Figure 1.) This means
that
\bea
C_c^k | \psi \rangle &  \approx & | \psi \rangle
\nonumber \\
C_c^{k'} | \psi \rangle  & \approx & 0 \ \ {\rm for } \ \ k' \ne k
\eea
and it follows that $D_{kk'} \approx 0 $. The key time scales here
are the classical arrival time for the centre of the packet,
\beq
t_a = \frac { m q_0 } { | p_0 | }
\eeq
and the Zeno time
\beq
t_z = \frac { m \sigma} { | p_0 | }
\eeq
(which is also approximately equal to $ 1 / (\Delta H_0) $). The Zeno time
is the time taken for the wave packet to move a distance equal to it spatial
width $\sigma$, or equivalently, it is the size of the packet's ``temporal
imprint'' at the origin. Therefore, the above approximations work if,
firstly,
\beq
t_z \ \ll \ \Delta
\eeq
and secondly, if the classical arrival time $t_a$ lies inside
the interval $[t_k, t_{k+1}]$ and is at least one or two Zeno times
away from the boundaries.

It is easy to see that
similar conclusions hold for superpositions of initial states of the
form Eq.(\ref{7.33}) as long as they are approximately orthogonal.
Loosely, this is because under the above conditions,
the class operators do not disturb the states and the only non-zero
components of the off-diagonal terms of the decoherence functional
will be proportional to the overlap of pairs of initial wave packets,
so will be approximately zero. (See Figure 3.) More general, non-orthogonal superpositions
may, however, produce backflow, so there may be no decoherence.

\subsection{Checking Decoherence for a Detailed Model}

Decoherence starts to become lost as the the size $ \Delta $ of the time intervals
$ [t_k, t_{k+1}]$ is reduced to close to the Zeno time. This is because
the wave packet will split into parts that cross during
different time intervals and the effect of diffraction in
time mentioned above \cite{diff} will cause these different parts
to be non-orthogonal. These effects will be explored in more detail
elsewhere \cite{HaYe}. Here, we give a more detailed calculation
to check for decoherence.

For simplicity we work with the simple case considered in Subsection 7D above.
We take the initial state to be the wave packet
Eq.(\ref{7.33}) and we note that the decoherence functional Eq.(\ref{7.21})
satisfies
\beq
| D |^2 \le 2 d_m^2
\eeq
with $d_m^2$ given by Eq.(\ref{7.30}) with $k=1$, $j=2$.
We need some probabilities
to compare this with. We have that
\beq
| D |^2 \le \langle C^2 \rangle \langle (1 - C)^2 \rangle
\eeq
The interesting case is that in which the crossing probability $ \langle C^2 \rangle$
is somewhat less than $1$, less than about $1/2$, say, in which case the non-crossing
probability $ \langle (1-C)^2 \rangle $ will be of order $1$. It is therefore
sufficient to compare $d_m^2$ with $ \langle C^2 \rangle $. Now note that
\bea
\langle C^2 \rangle &=& \langle \left( \theta ( \hat x_1 ) - \theta ( \hat x_2 ) \right)^2 \rangle
\nonumber \\
&=& \langle \theta ( \hat x_1 ) \theta ( -\hat x_2 )  \theta (  \hat x_1 ) \rangle
+ \langle \theta ( \hat x_1 ) \theta ( -\hat x_2 )  \theta (  - \hat x_1 ) \rangle
+ \langle \theta ( \hat x_2 )  \theta (  - \hat x_1 ) \rangle
\eea
This means that $\langle C^2 \rangle $ is in fact equal to the probability
\beq
p_{12} = \langle \theta ( \hat x_1 ) \theta ( -\hat x_2 )  \theta (  \hat x_1 ) \rangle
\label{7.39b}
\eeq
up to terms which vanish when $D = 0$. This is useful since it is now identical in form
to the expression for $d_m^2$ and our goal is to show that
\beq
d_m^2 \ll p_{12}
\eeq

It is useful to work in the Wigner representation \cite{Wig}, defined, for a state $\rho(x,y)$
by
\beq
W(p,q) = { 1 \over 2 \pi } \int d \xi \ e^{- i  p \xi}
\ \rho( q + \half \xi, q - \half \xi)
\eeq
The probabilities $p_{12}$ and $d_m^2 $ are then given by
\bea
p_{12} &=& 2 \pi \int dp dq \ W_{12} (p,q) \ W_0 (p,q,t_1)
\label{7.39}
\\
d_m^2 &=& 2 \pi \int dp dq \ W_ D (p,q) \ W_0 (p,q,t_1 )
\label{7.40}
\eea
Here, $  W_0 (p,q,t_1 ) $ is the Wigner function of the initial state,
evolved in time to $t_1$,
\beq
 W_0 (p,q,t_1 ) = \frac {1} {\pi} \exp \left( - \frac { (q - q_0 - p_0 t_1 / m)^2 } { 2 \sigma^2}
 - 2 {\sigma^2}   ( p - p_0 )^2 \right)
\label{7.41}
\eeq
The objects $W_{P}$ and $W_D$ are the Wigner transforms of the $\theta$-function
combinations appearing in Eqs.(\ref{7.30}), (\ref{7.39b}) and are given by
\bea
W_{P} (p,q) &=& \frac {1} {2 \pi^2} \theta (q) \int_{u(p,q)}^\infty dy \ \frac { \sin y } {y}
\label{7.42}
\\
W_D (p,q) &=& \frac {1} {2 \pi^2} \theta (- q)  \int_{u(p,q)}^\infty dy \ \frac { \sin y } {y}
\label{7.43}
\eea
where
\beq
u(p,q) = 2 q \left( p + \frac {m q } {(t_2 - t_1)} \right)
\eeq
We see that the only difference between the expressions for $p_{12}$
and $d_m^2$ is in the
sign in the $\theta$-functions.

The integral
\beq
f(u) = \int_{u}^\infty dy \ \frac { \sin y } {y}
\eeq
may be expressed in terms of the Sine integral function ${\rm Si} (x)$,
\beq
f(u) = \frac {\pi } {2} - {\rm Si} (u)
\eeq
but its properties are not hard to see directly. For large negative $u$,
$ f(u) \approx \pi $, at $u=0$, $f(0) = \pi / 2 $, and for large
positive $u$, $f(u)$ goes to zero, oscillating around $ 1/u$.
(See Figure 4.)

We now compare the size of $p_{12}$ and $d_m^2$.We assume that the wave packet
is spatially broad, so $\sigma$ is large and the Wigner function Eq.(\ref{7.41})
is therefore concentrated strongly about $ p = p_0 < 0 $. We therefore
integrate out $p$ and set $ p = p_0$ throughout.
The most important case to check is that in which the wave packet is reasonably
evenly divided between $ x>0$ and $x<0$ at time $t_1$ so that both $p_{12}$
and $d_m^2$ have a chance of being reasonably large. This means that
$ q_0 + p_0 t_1 / m $ should be close to zero (to within a few widths $ \sigma$),
so for simplicity we take it to be exactly zero.

With these simplifications, we have
\beq
d_m^2 = \frac {1 } {(2 \pi^3 \sigma^2)^{1/2}} \int_{-\infty}^0 dq \ \exp \left( - \frac {q^2 } {2 \sigma^2} \right)
\ f ( u (p_0, q) )
\label{7.47}
\eeq
Here, since $q \le 0$ and $p_0 < 0 $, we have $ u(p_0, q) \ge 0 $. We can evaluate
this expression by examining the comparative effects of  $f(u)$ and the exponential
term. From the plot
of $f(u)$ (see Figure 4), we see that it drops to zero at $ u=u_0$ (which is of order $1$)
and oscillates rapidly
around zero for $u>u_0$, so we expect the integral to be dominated by values of
$q$ for which $ 0 \le u \le u_0$. The value $u=u_0$ corresponds
to $q=q_0$ where
\beq
q_0 = - \frac { | p_0 | \Delta } { 2 m } \left( \left[ 1 + \frac {u_0} { E_0 \Delta } \right]^{1/2}
 -1 \right)
\eeq
where $E_0 = p_0^2 / 2m $. In the complex potential calculations, we have assumed
that $ E_0 \gg V_0 $ and we also assumed that all timescales are much greater than $ 1 / V_0 $,
and these together imply that $ E_0 \Delta \gg 1 $. We may therefore expand the square root
to leading order and obtain
\beq
q_0 \approx - \frac { u_0 } { 2 |p_0 |}
\eeq
We have assumed that the wave packet is sufficiently broad that $ \sigma p_0 \gg 1 $,
and this means that
\beq
|q_0| \ll \sigma
\eeq
This in turn means that $f(u)$ is significantly different from zero
only in the range $ | q | \ll \sigma $, and most importantly, in this range, the exponential term
in Eq.(\ref{7.47}) is approximately constant. We may therefore evaluate Eq.(\ref{7.47})
by ignoring the exponential term, integrating from $ 0 $ to $ q_0$ and
approximating $f(u)$ as
\beq
f(u) \approx \frac { \pi } {2} - u + O (u^3)
\eeq
We thus obtain the simple result,
\beq
d_m^2 \approx \frac {1 } {(2 \pi^3 )^{1/2}} \ \frac {1} { |p_0| \sigma} \ll 1
\eeq

In the expression for $p_{12}$, there is a key difference in that
$q>0$ which means that $ u(p_0,q)$ can be positive or negative.
Introducing
\beq
q_z  = \frac {| p_0 |} {m} \Delta  = \sigma \frac { \Delta  } {t_z}
\eeq
(where $t_z$ is the Zeno time) we see that $ u < 0 $ for $ q< q_z $ and $ u> 0 $ for $ q > q_z $.
We therefore have
\bea
p_{12} &=& \frac {1 } {(2 \pi^3 \sigma^2)^{1/2}} \int_0^{q_z} dq
\ \exp \left( - \frac { q^2 } { 2 \sigma^2}
\right) \ f ( u (p_0, q) )
\nonumber \\
& &  \ \ \ \ \ + \frac {1 } {(2 \pi^3 \sigma^2)^{1/2}} \int_{q_z}^{\infty} dq \ \exp \left( - \frac { q^2 } { 2 \sigma^2}
\right) \ f ( u (p_0, q) )
\label{7.49}
\eea
Here, $ f(u(p_0,q)) \approx \pi $ in the first term, differing from this value only in
a region of size $1/p_0$ close to $ q=0$. In the second term $f(u)$ will
tend to be small except for a small region of size $ 1 / p_0 $ around the origin.

If $q_z \gg \sigma $, (that is, $ \Delta \gg t_z$), the second term in $p_{12}$
is exponentially suppressed and in the first
term the integration range is effectively $ 0$ to $ \infty $, so we obtain
\beq
p_{12} \approx \frac {1 } {2}
\eeq
This is the expected result, since under the above assumptions on the wave
packet, half of it will cross $x=0$ if the time interval is sufficiently
large. Clearly $ p_{12} \gg\ d_m^2 $ in this case so there is decoherence.

If $q_z \ll \sigma $, the first term in $p_{12}$ is of order $ q_z / \sigma $ and the second
of order $ 1 / ( |p_0| \sigma ) $, the same order of magnitude as $d_m^2$. Hence
in this case we have decoherence if
\beq
\frac { \Delta } { t_z } \gg \frac { 1 } { |p_0| \sigma}
\eeq
Since the right-hand side is already $ \ll  1$, this is easily
satisfied even for time intervals whose size is $\Delta $ is smaller
than the Zeno time. In fact this condition is equivalent to the condition
$ E_0 \Delta \gg 1 $, which is satisfied by the assumptions of the
complex potential model, as stated above.

In brief, we therefore get decoherence of histories for a single
wave packet under a wide variety of circumstances.

\section{Summary and Conclusions}

This paper was initially motivated by a desire to analyse the arrival time problem
using the decoherent histories approach to quantum theory. But along the way we have
reconsidered and derived a number of other useful related results.

We considered the arrival time problem using a complex potential to kill paths entering
$x<0$. In Section 2 we gave a classical analysis of the problem. We derived a
result of the expected form exposing the resolution function as an essentially
classical effect summarizing the role of the complex potential. We also showed
that coarse graining over time scales much greater than $1/ V_0$ produces a formula
for the arrival time of expect form and which is independent of the complex potential.
This is an important result for the rest of the paper.

After reviewing the path decomposition expansion in Section 3, we used it in Section 4 to
derive the standard results for stationary scattering states. This is simpler
and more direct than the usual method, involving matching eigenfunctions. (In a sense
this cumbersome machinery is already encoded in the PDX). In Section 5, we used the
PDX to rederive the standard form of the arrival time distribution with a complex potential,
in the limit of weak potential. The form of this calculation turned out to be useful for
the subsequent work on the decoherent histories approach.

In Section 6, we considered the decoherent histories analysis for the simple case
of a particle crossing or not crossing $x=0$ during a large time interval $[0,\tau]$.
We found the simple and expected result that the histories are decoherent as long
as reflection by the complex potential is negligible. The resultant probabilities
are consistent with the standard result for the arrival time.

The main part of the decoherent histories analysis was given in Section 7, where we first
derived the class operators described crossing $x=0$ for an arbitrary set of small
time intervals. Here we obtained our most important result: the crossing class operator
Eq.(\ref{7.15}) for timescales much greater than $1/V_0$. This form of the class operator
gives an immediate connection with the standard result for probabilities when there is
decoherence. Indeed, one may have {\it guessed} the form of the class operator from
the standard form of the probabilities, and this is pursued in another paper \cite{HaYe2}.
However, it is also gratifying that it can be derived in some detail using the complex
potential approach used here.

To assign probabilities, the decoherence functional must be diagonal and we considered
this condition. We found a variety of states for which there is decoherence, under
certain more detailed conditions, which we discussed.

We also noted an interesting and important relationship between decoherence and backflow:
If there is decoherence, the probabilities for crossing must be positive so there cannot
be any backflow. If there is no decoherence, the integrated current may still be positive,
but one can say that if there is backflow there will definitely be no decoherence.
This means that the decoherent histories approach brings something genuinely new to
the arrival time problem: it establishes the conditions under which probabilities can be
assigned and in particular forbids the assignment of probabilities in cases where there
is backflow.

\section{Acknowledgements}

We are grateful to Gonzalo Muga and Larry Schulman for useful
discussions. We are also very grateful to Chris Dewdney for supplying
a computer programme representing the evolution of wave packets, which was very useful
at an early stage in this work. JJH acknowledges the hospitality of
the Max Planck Institute in Dresden at which some of this work was
carried out during the Advanced Study Group, ``Time: Quantum and Statistical Mechanics Aspects",
August 2008.

\appendix

\section{Some Properties of the Current}

We have derived the expression
\beq
p (0, T) = \int^{T}_{0} dt \ J(t)
\eeq
as the approximate probability for crossing the origin during the time interval $[0,T]$, where
$J(t)$ is the usual quantum-mechanical current. The current itself is not necessarily positive
due to backflow. Here we explore the possibility that averaging it over time might improve
the situation. On the one hand, the results of Bracken and Melloy \cite{BrMe} show that
there is always {\it some} state for which $p(0,T)$ defined above is negative, for any $T$.
On the other hand, for a {\it given} state, one might hope that $p(0,T)$ will be positive
for sufficiently large $T$. Here we give a brief argument for this, which also makes
contact with the negativity of the Wigner function.

The current can be written in terms of the Wigner function $ W(p,x)$ as
\beq
J(t)= - \int dp \ \frac{p}{m}\  W(p,0,t).
\eeq
The Wigner function evolves freely according to $W(p,x,t)=W(p,x-pt/m,0)$. We assume
it has support only on negative momentum states, with average momentum $p_0 <0$
and momentum width $\sigma_p$.

Consider a time interval $0<t < T$ over which backflow occurs. It is clear that in order
for this to occur the Wigner function must be negative for at least some of this interval.
We can write,
\bea
p(0,T) &=& - \int^{T}_{0}dt\ \int dp \ \frac{p}{m}  \ W(p,-pt/m,0)
\nonumber \\
&=&  \int dp \ \int_0^{|p|T/m}  dx\   \ W(p,x,0)
\eea
So $p(0,T)$ is given by the average of the Wigner function over a region of
phase space. We now recall a standard property of the Wigner function
which is that, broadly speaking, it will tend to be positive when averaged
over a region of phase space of size greater than order $1$ (in the units used
here where $\hbar = 1 $).
This region is of size of order $ | p_0 | T / m $ in the $x$-direction but
infinite in the $p$-direction. However, the Wigner function has momentum
spread $\sigma_p$, so the effective size averaged over is $\sigma_p p_{0} T/m$
which is approximately the same as $\Delta H\ T$. This means that we expect that
$p(0,T)$ will be positive as long as
\beq
T  \ > \ \frac {1} { \Delta H}
\eeq
Hence, as expected, the integrated current will be positive for $T$ sufficiently
large and the key timescale is the Zeno time. This heuristic argument
will be revisited in more detail elsewhere.

\bibliography{apssamp}

\vfil \eject

\epsfbox{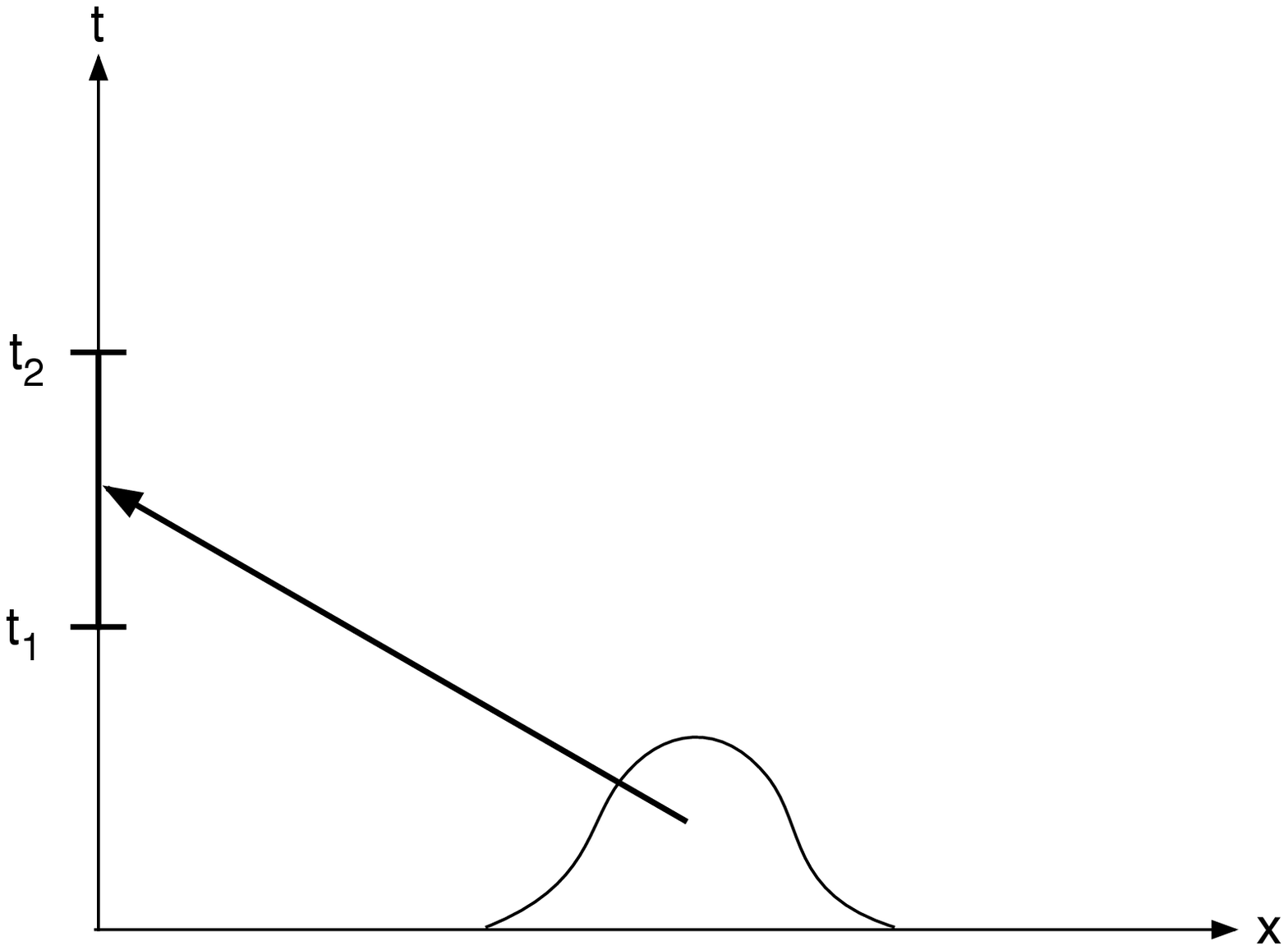}

\noindent{\bf Figure 1.} The quantum arrival time problem. We prepare an initial state
localized entirely in $x>0$ and consisting entirely of negative momenta. What is
the probability that the particle crosses the origin during the time interval $[t_1,t_2]$?

\epsfbox{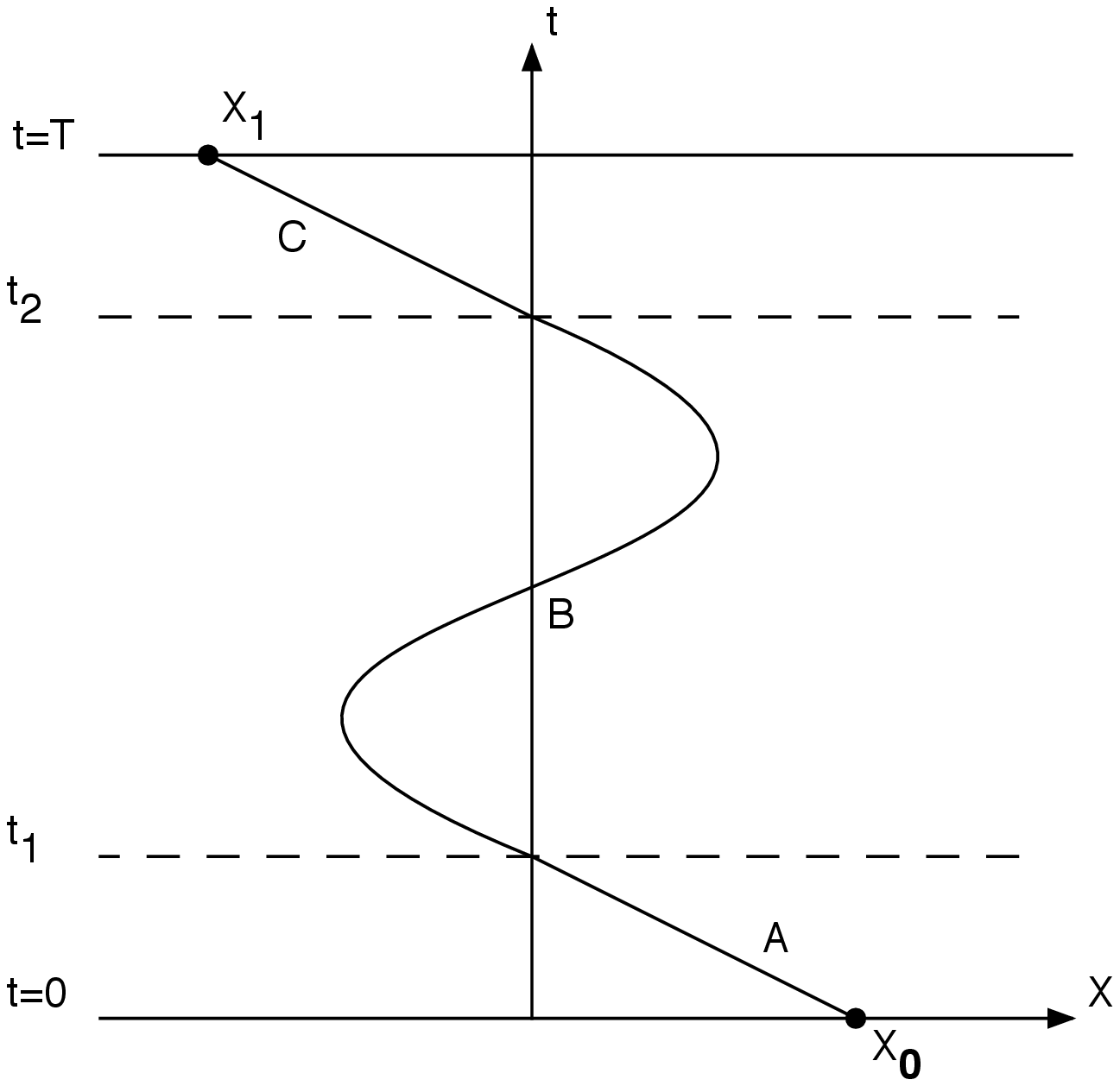}

\noindent{\bf Figure 2.} The path decomposition expansion. Any path from $ x_0 > 0 $
at $t=0$ to a final point $x_1 < 0 $  at $t=T$ has a first crossing of $x=0$ at $t_1$ and
a last crossing at $t_2$. The propagator from $(x_0,0)$ to $(x_1, T)$ may be decomposed
into three parts: (A) restricted propagation entirely in $x>0$, (B) free propagation
starting and ending on $x=0$, and (C) restricted propagation entirely in $x<0$.
The corresponding path decomposition expansion formula is given in Eq.(\ref{16a}).

\epsfbox{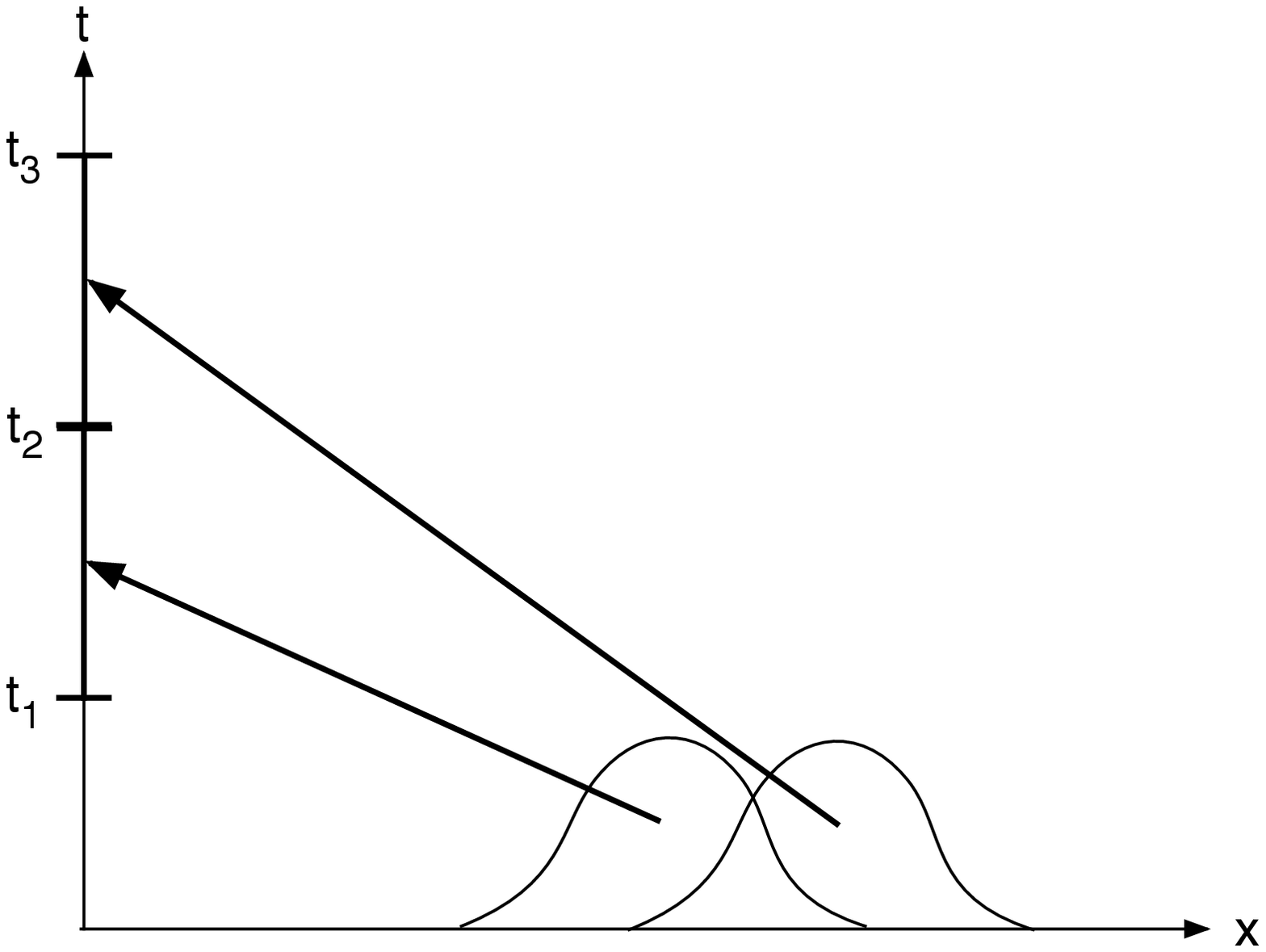}

\noindent{\bf Figure 3.} An initial state consisting of a superposition of wave packets
may have significant crossings in at least two different time intervals. If the initial
packets are orthogonal, and the time intervals are sufficiently large (greater than the
Zeno time), the packets will remain orthogonal after passing through the time intervals
and the corresponding histories will be decoherent.

\vfil\eject

\epsfbox{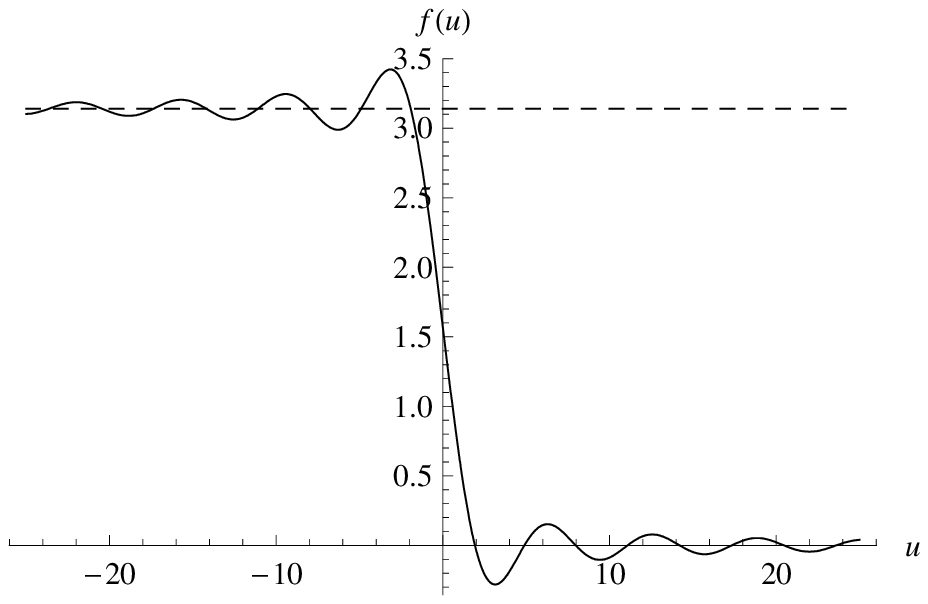}

\noindent{\bf Figure 4.} The function $f(u)$. It oscillates around zero for $u \gg 0$
and oscillates around $ \pi $ for $ u \ll 0 $. As a function of $q$, $f(u(p_0,q))$
differs from $0$ or $ \pi$ only in a region of size $1/|p_0|$ around $q=0$.

\end{document}